\documentclass[journal=jacsat,manuscript=article]{achemso}

\usepackage[version=3]{mhchem} 
\usepackage[dvipsnames]{xcolor}
\usepackage{bm}
\usepackage{threeparttable}
\usepackage{amsmath,amsfonts,amssymb}
\usepackage{booktabs}
\usepackage{graphicx,caption}
\usepackage{soul}
\usepackage{array}
\usepackage{lscape}

\author{Apurba Nandi}
\email{apurba.nandi@uni.lu}
\affiliation{Department of Physics and Materials Science, University of Luxembourg, L-1511, Luxembourg City, Luxembourg.}

\author{Priyanka Pandey}
\affiliation{Department of Chemistry and Cherry L. Emerson Center for Scientific Computation, Emory University, Atlanta, Georgia 30322, U.S.A.}
\author{Paul L. Houston}
\affiliation{Department of Chemistry and Chemical Biology, Cornell University, Ithaca, New York
14853, U.S.A. and Department of Chemistry and Biochemistry, Georgia Institute of
Technology, Atlanta, Georgia 30332, U.S.A}
\author{Chen Qu}
\affiliation{Independent Researcher, Toronto, Ontario M9B0E3, Canada}

\author{Qi Yu}
\affiliation{Department of Chemistry, Fudan University, Shanghai, 200438, P. R. China }

\author{Riccardo Conte}
\affiliation{Dipartimento di Chimica, Universit\`{a} degli Studi di Milano, via Golgi 19, 20133 Milano, Italy}
\author{Alexandre Tkatchenko}
\email{alexandre.tkatchenko@uni.lu}
\affiliation{Department of Physics and Materials Science, University of Luxembourg, L-1511, Luxembourg City, Luxembourg.}
\author{Joel M. Bowman}
\email{jmbowma@emory.edu}
\affiliation{Department of Chemistry and Cherry L. Emerson Center for Scientific Computation, Emory University, Atlanta, Georgia 30322, U.S.A.}

\title[] { $\Delta$-Machine Learning to Elevate DFT-based Potentials and a Force Field to the CCSD(T) Level Illustrated for Ethanol}

\abbreviations{PES}
\keywords{American Chemical Society, \LaTeX}

\begin{document}

\begin{tocentry}
\includegraphics[width=1.75in]{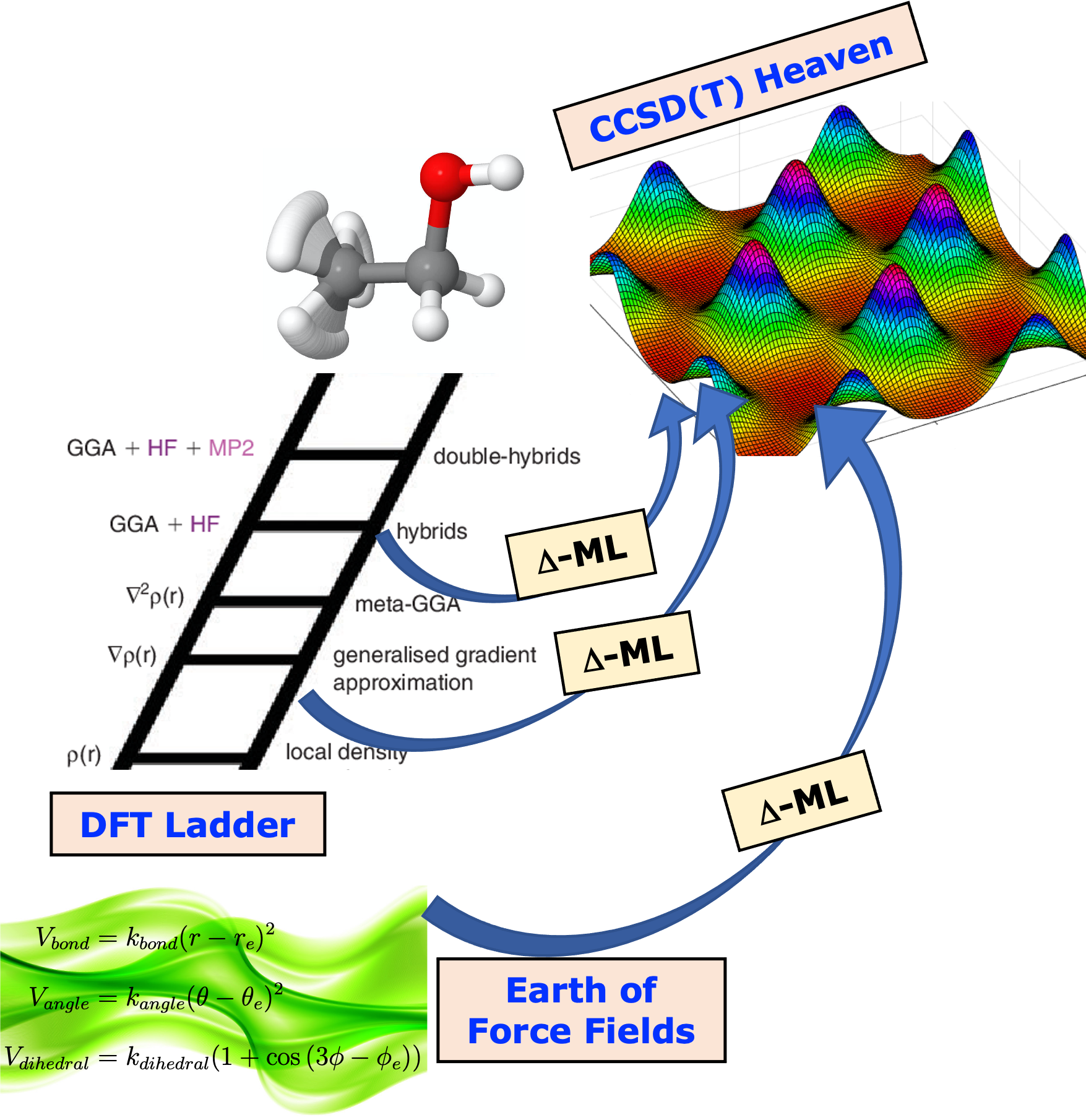}
\end{tocentry}

\newpage
\begin{abstract}

Progress in machine learning has facilitated the development of potentials that offer both the accuracy of first-principles techniques and vast increases in the speed of evaluation. Recently,``$\Delta$-machine learning'' has been used to elevate the quality of a potential energy surface (PES) based on low-level, e.g., density functional theory (DFT) energies and gradients to close to the gold-standard coupled cluster level of accuracy. We have demonstrated the success of this approach for molecules, ranging in size from \ce{H3O+} to 15-atom acetyl-acetone and tropolone. These were all done using the B3LYP functional.  Here we investigate the generality of this approach for the PBE, M06, M06-2X, and PBE0+MBD functionals, using ethanol as the example molecule. Linear regression with permutationally invariant polynomials is used to fit both low-level and correction PESs. These PESs are employed for standard RMSE analysis for training and test datasets, and then general fidelity tests such as energetics of stationary points, normal mode frequencies, and torsional potentials are examined. We achieve similar improvements in all cases. Interestingly, we obtained significant improvement over DFT gradients where coupled cluster gradients were not used to correct the low-level PES.
Finally, we present some results for correcting a recent molecular mechanics force field for ethanol and comment on the possible generality of this approach.

\end{abstract}
\newpage
\section{Introduction}

Developing high-dimensional, \textit{ab initio}-based potential energy surfaces (PESs) is an active area of theoretical and
computational research. Major progress has been made in using and developing machine learning (ML) approaches for PESs with more than five atoms, based on fitting thousands of CCSD(T) energies~\cite{bowman11, ARPC2018, Fu18, guo20,robo20} or forces~\cite{Tkatch2018, Tkatch19}. Some of these ML approaches have been using permutationally invariant polynomials (PIPs) or PIPs as inputs to neural network software.\cite{bowman11, ARPC2018, Fu18, guo20, robo20}  
Of course, there are numerous other ML methods.  It is perhaps of interest and relevance to this paper that the precision of a PIP PES for ethanol was shown to be as good as the best performing ML methods and to be substantially faster (factors of 10 or more)\cite{Bowman_reverse2022} than all the ML methods considered, i.e., GAP-SOAP,\cite{GP-2015-1} ANI,\cite{AN1} DPMD,\cite{dpmd2018} sGDML,\cite{Tkatch2018, Tkatch19} PhysNet,\cite{PhysNet} KREG,\cite{KREG} and pKREG \cite{pKREG}.  The dataset for that method was generated using 500 K direct dynamics based on the PBE0 functional.  CCSD(T) datasets for larger molecules are rare, and the 10-atom formic acid dimer is one prominent recent example; PESs for this dimer have been reported using PIPs\cite{Qu2016} and later an atom-centered high-dimensional NN.\cite{fad2022} Complex reactive potentials for 6 and 7-atom chemical reactions, which are fitted to tens of thousands or even hundred thousand CCSD(T) energies, have been reported.\cite{Furoam20, HCH3OH}. The PIP-based automated ROBOSURFER software\cite{robo20} has been applied to develop a number of complex PESs for 9-atom chemical reactions.\cite{cazko9atom,czako2024} 

There is a major bottleneck to use this level of theory with the increase in molecular size. It is the steep scaling of CCSD(T) calculations of order $\sim N^7$, $N$ being the number of basis functions.  

Correcting \textit{ab initio}-based potential energy surfaces has been a long-standing goal of computational chemistry.  In one approach, dating back 30 years, a correction potential is added to an existing PES, and parameters of the correction potential are optimized by matching ro-vibrational energies to experiment.\cite{Light1986, WU1996, SKOKOV99}  This approach relies on being able to calculate exact ro-vibrational energies to make the comparison with the experiment robust. Thus, it has only been applied to triatomic molecules and it is limited to these and possibly tetratomics.  Another approach is to modify an existing potential using scaling methods that go under the heading of  ``morphing".\cite{gbhcnmorph, gbscaling, meuwlymorph} 

More recent approaches using machine learning aim to bring a PES, based on a low-level of electronic theory, typically density functional theory, to a higher level such as coupled cluster (CC) theory. In consideration of larger molecules and clusters, where high-level methods are prohibitively expensive, the motivation for doing this is clear.  There are two classes of such approaches, one is ``$\Delta$-machine learning" ($\Delta$-ML) and the other is ``transfer learning".\cite{TL_ieee} $\Delta$-ML, which is of direct relevance to the present paper, seeks to add a correction to a property obtained using an efficient and thus perforce low-level \textit{ab initio} theory.\cite{Lilienfeld15, Lilienfeld19, Tkatch19, Tkatch2018, Stohr2020} A hierarchical $\Delta$-ML method using multiple quantum chemistry methods has been applied to a \ce{CH3Cl} PES\cite{Csanyi_DeltaML}.   In this sense, the approach is related, in spirit at least, to the correction potential approach mentioned above, when the property is the PES.  However, it is applicable to much larger molecules. 

Notable recent applications of transfer learning have been reported by Meuwly and co-workers to improve neural network PESs for malonaldehyde, acetoacetaldehyde, and acetylacetone.\cite{meuwly2022}

We recently suggested a single-step $\Delta$-ML approach to bring a DFT-based PES to the CCSD(T) level.\cite{Nandi_Bowman_DeltaML} The simple equation describing this is given below, Eq. \ref{eq:1}.  Subsequently, this method has been applied successfully to acetylacetone \cite{acac2021,AcAc_MTA,NANDI_AICHEM}, ethanol\cite{nandi2022quantum,ethanolMM}, tropolone\cite{Nandi2023JACS}, and the formic acid-ammonia dimer\cite{fanh32024}. Additionally, the approach has also been proposed to correct many-body force fields.\cite{JCTCPers2023} In all these examples, the B3LYP functional was used to obtain the DFT-based PIP PES.  

Considering the success of the $\Delta$-ML method with the B3LYP functional,\cite{LWP,becke93} it is both interesting and significant to explore whether this straightforward approach can be extended to other functionals and molecular mechanics, including ·`classical'‘ force fields (FFs).
There is a vast literature on molecular mechanics force fields, and the reader is directed to a recent and relevant (vide infra) paper that surveys this field.\cite{meuwlyFFDelta} While these FFs, which are heavily semi or totally empirical, have made an enormous impact in biomolecular simulations, there is strong motivation to progress from these.  Broadly put, there are two approaches that can be undertaken.  One is to replace these FFs with strictly ML FFs, based on electronic structure energies and forces for the covalent and non-covalent interactions, and sophisticated treatments of long-range interactions. This is a major challenge for an ML method that aims to deal with hundreds of atoms in a single step. A recent example of this approach by Tkatchenko, M{\"u}ller and co-workers can be found in ref. \citenum{sciadv2024}. Of course, invoking the ``no free lunch" axiom, this approach is far more demanding in computational effort compared to a classical FF.  A second approach is to correct a classical force field. There have been several papers along these lines including one from this group aimed at correcting a sophisticated classical FF for water, by correcting the short-range 2-b, 3-b, and even 4-b interactions.\cite{JCTCPers2023}  However, while water is essential for life it is not a biomolecule. Other similar approaches, focused on correcting the short-range interactions have also appeared recently \cite{D3SC02581K}.  While these approaches may be less computationally demanding than a full ML approach, they are still far more demanding than biomolecular FFs.

A variation of the second approach, which is our focus, is to continue to use the classical FF expression for the potential, i.e., harmonic bond stretches and bends, periodic torsional potentials, plus simple 2-b non-covalent interactions and long-range electrostatics and to add a computationally efficient ML correction.  To facilitate the goal of efficiency, the ML correction can be applied after making a correction to some terms, at least, in the classical FF are corrected using \textit{ab initio} electronic energies. Recently, Meuwly and co-workers,\cite{Meuwly2023,meuwlyFFDelta} investigated correcting the CHARMM classical force field for specific examples.  An earlier, but still recent, example of this approach used atomic force matching (AFM), using MP2 theory, to determine classical FF intramolecular terms of ethanol plus the 2-b intermolecular interaction between an ethanol and water molecules.\cite{rogers2020accurate} Here, we use this AFM-corrected FF for ethanol to investigate our computationally efficient $\Delta$-ML approach, which substantially improve several key properties of AFM-corrected FF. Most notably, it addresses the harmoinc normal-mode frequencies, which are greatly overestimated by this FF for all but the lowest several normal modes.

The paper is organized as follows.  A brief review of the $\Delta$-ML approach is provided, along with the essentials of the highly efficient ML linear-regression approach we use with permutationally invariant polynomials. Results and discussion follow, including remarks on the extension of the $\Delta$-ML PIP approach to much larger molecules. 

\section{Theory}
\subsection{$\Delta$-ML approach}
The $\Delta$-ML approach is given by the equation
\begin{equation} 
\label{eq:1}
    V_{LL{\rightarrow}CC}=V_{LL}+\Delta{V_{CC-LL}},
\end{equation}
where $V_{LL{\rightarrow}CC}$ is the corrected PES,  $V_{LL}$ is a PES fit to low-level DFT electronic data, and $\Delta{V_{CC-LL}}$ is the correction PES based on high-level coupled cluster energies. To investigate the efficacy of the $\Delta$-ML approach, four widely used functionals are employed here, M06\cite{MO6} and M06-2X\cite{MO6,M06_2X} functionals with the 6-311+G(d,p) basis, PBE\cite{PBE96} with the def2-SVP basis and PBE0\cite{PBE0_1,PBE0_2} including many-body dispersion (MBD)\cite{mbd2012} with the ``intermediate'' basis setting.\cite{FHI_aims} Additionally, we also replace $V_{LL}$ with a classical force field.  

Previously, it is noted that the difference between CCSD(T) and DFT energies, $\Delta{V_{CC-LL}}$, does not vary as strongly as $V_{LL}$ with respect to the nuclear configurations and therefore, a small number of high-level electronic energies is adequate to fit the correction PES.  

It is not clear if this observation applies, at least semi-quantitatively, for classical force fields.  In this case, the differences can be much larger, as expected and indeed verified here for ethanol. We investigate this using a previous dataset of 2319 CCSD(T)-F12a/aug-cc-pVDZ electronic energies.\cite{nandi2022quantum}

The permutationally invariant polynomial (PIP) approach is used to fit both the $V_{LL}$ and $\Delta{V_{CC-LL}}$ PESs.
The theory of permutationally invariant polynomial is well established and has been presented in several review articles.\cite{Braams2009, Bowman2010, Xie10, bowman11, ARPC2018} In terms of a PIP basis, the potential energy, $V$, can be written in compact form as

\begin{equation}
\label{eq:2}
V(\bm{x})= \sum_{i=1}^{n_p} c_i p_i(\bm{y}),
\end{equation}
where $c_i$ are linear coefficients, $p_i$ are PIPs, $n_p$ is the total number of polynomials for a given maximum polynomial order, and $\bm{y}$ are the collection of Morse variables.  For example, $y_{\alpha \beta}$ is given by $\exp(-r_{\alpha \beta}/\lambda)$, where $r_{\alpha \beta}$ is the internuclear distance between atoms $\alpha$ and $\beta$. The range (hyper)parameter, $\lambda$, is typically 2-3 bohr. The linear coefficients are obtained using standard least squares methods for large data sets of electronic energies (and for large molecules' gradients as well) at scattered geometries.

\subsection{The Ethanol Force Field}
\begin{figure}[htbp!]
    \includegraphics[width=0.7\columnwidth]{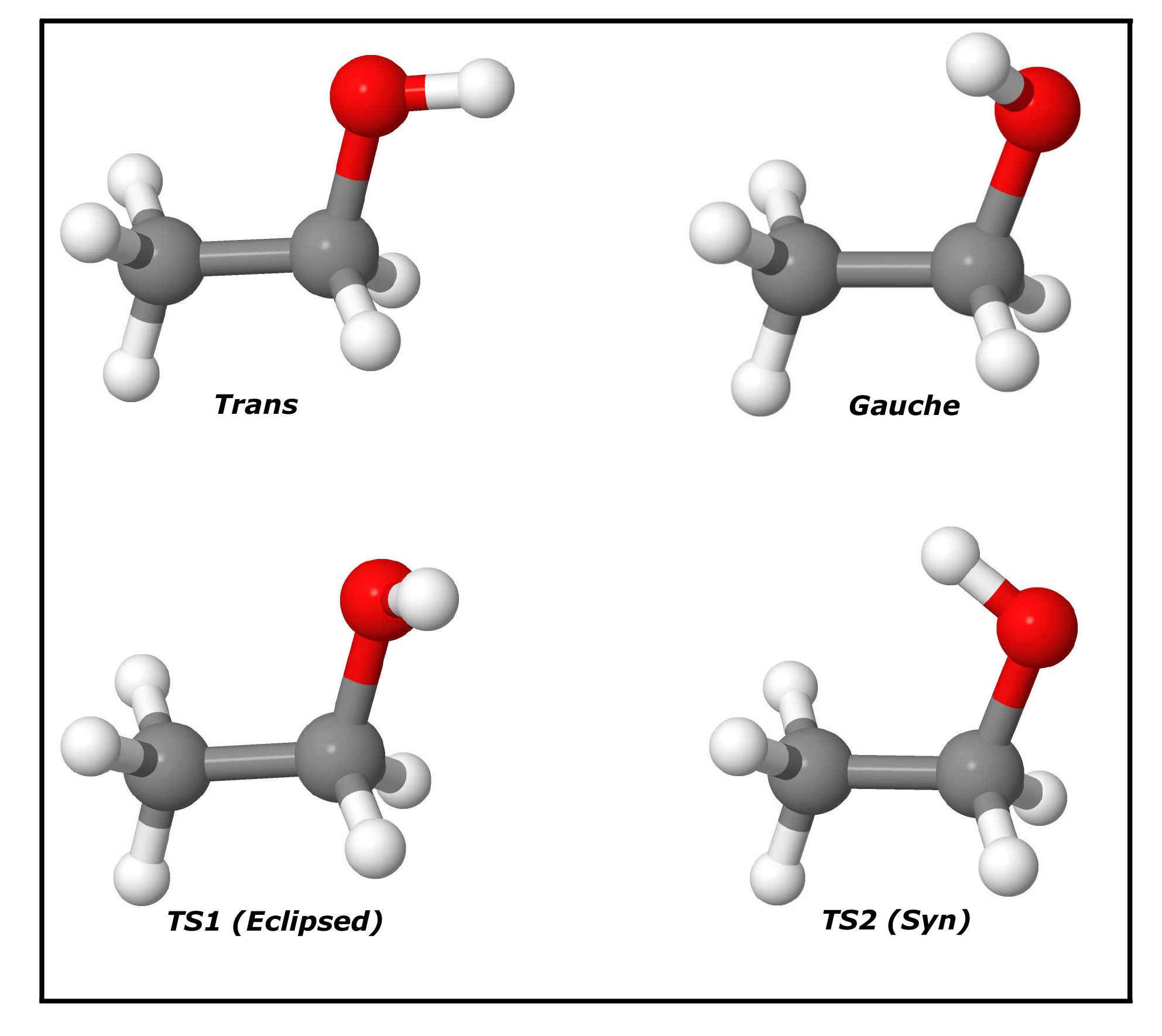}
    \caption{Optimized geometry of $trans$ and $gauche$ conformers of ethanol and their two isomerization TSs at CCSD(T)-F12a/aug-cc-pVDZ level. Reproduced from  Nandi, A.; Conte, R.; Qu, C.; Houston, P. L.; Yu, Q.; Bowman, J. M. Quantum Calculations on a New CCSD(T) Machine-Learned Potential Energy Surface Reveal the Leaky Nature of Gas-Phase Trans and Gauche Ethanol Conformers. \emph{J. Chem. Theory Comput.} {\bf 2022}, 18, 5527-5538. Copyright 2022 American Chemical Society.}
        \label{fig:geom}
\end{figure}

Figure \ref{fig:geom} shows conformations of $trans$ and $gauche$-ethanol and two saddle point transition states.  These are from electronic structure calculations at the CCSD(T)-F12a/aug-cc-pVDZ level. 

The molecular mechanics force field we consider is the recent one that was corrected using force matching MP2 gradients computed with triple-zeta-quality basis sets using the Adaptive Force Matching method~\cite{rogers2020accurate}. The potential form of the force field includes intramolecular interaction terms for interactions of atoms that are linked by molecular bonds.

\begin{align}
     V_{FF}    =  V_{bond} + V_{angle} + V_{dihedral} 
     \label{eqn:ff_potn}
\end{align}
where $V_{bond}$ and $V_{angle}$ are modeled by the quadratic energy functions, corresponding to the oscillations about an equilibrium bond length and bond angle, based on the Harmonic approximation, and $V_{dihedral}$ is modeled by the cosine function, corresponding to the torsional rotation of four atoms about a central bond.
\begin{align}
    \label{eqn:bond_ha}
    V_{bond} &= k_{bond}(r - r_e )^2\\ 
    V_{angle} &= k_{angle}(\theta -\theta_e )^2 \\
    V_{dihedral} &= k_{dihedral}(1 + \cos{(3\phi - \phi_e)})
\end{align}
The value of fitting parameter $k_{bond}$, $k_{angle}$, $k_{dihedral}$ as well as the equilibrium bond lengths and angles are taken from Ref. \citenum{rogers2020accurate}. 

\section{Computational Details}
We use the dataset from our recently reported ``MDQM21'' dataset\cite{Bowman_reverse2022}, which includes a total of  11000 energies and their corresponding gradients generated from \textit{ab initio} molecular dynamics (AIMD) simulations at B3LYP/6-311+G(d,p) level of theory. This dataset was partitioned into a training set of 8500 configurations and a test set of 2500 configurations. The same training and test dataset were used for single point energy and gradients computations at M06/6-311+G(d,p),\cite{MO6} M06-2X/6-311+G(d,p),\cite{MO6,M06_2X} and PBE/def2-SVP\cite{PBE96} level of theory using MOLPRO\cite{MOLPRO_brief} quantum chemistry package and at PBE0+MBD\cite{PBE0_1,PBE0_2,mbd2012} level of theory with ``intermediate'' basis setting using the FHI-aims electronic structure package.\cite{FHI_aims,FHI_aims_2}

\section{Results and discussion}
\subsection{$\Delta$-ML for DFT Functionals}
The low-level PES, V$_{LL}$, is fitted using a maximum polynomial order of 4 with permutational symmetry 321111, utilizing 8500 DFT data (all four different DFT functionals) points. This results in a total of 14752 PIPs in the fitting basis set. Testing was done on 2500 data points.
The root mean square (RMS) fitting errors for training and test data sets are shown in Table~\ref{tab:rmse_dft}.

\begin{table}[h]
    \centering
    \begin{tabular}{c|c|c|c|c|c}
    \hline  \hline
                 & PBE  & M06 & M06-2X & B3LYP & PBE0+MBD \\ [0.5ex] \hline  
    Training     &  45  &  79 &   47   &  40 & 40\\
    Test         &  56  &  82 &   57   &  51 & 51\\ [0.5ex]
        \hline \hline
    \end{tabular}
    \caption{The RMS Fitting Error (in cm$^{-1}$) of V$_{LL}$ for training and test data sets.  }
    \label{tab:rmse_dft}
\end{table}

Next, we train $\Delta$-correction PES on the difference between the CCSD(T) and DFT absolute energies of 2069 geometries and test the obtained surface on the remaining 250 geometries. To fit the  $\Delta V_{LL \rightarrow CC}$, we have used a maximum polynomial order of 2 with permutational symmetry 321111 for the training data set. This results in a basis size of 208 PIPs generated using our MSA software~\cite{msachen,msavideo}. The RMS training and test errors for the energies of correction PES are shown in Table~\ref{tab:rmse_deltapes}.

\begin{table}[h]
    \centering
    \begin{tabular}{c|c|c|c|c|c}
    \hline \hline
                 & PBE  & M06 & M06-2X & B3LYP & PBE0+MBD \\ [0.5ex] \hline    
    Training     &  67  &  53 &   32  &  28 & 26\\
    Test         &  90  &  61 &   40  &  30 &  30\\ [0.5ex]
        \hline \hline
    \end{tabular}
    \caption{The RMS Fitting Error (in cm$^{-1}$) of correction PESs $\Delta V_{CC-LL}$ for training and test data sets.}
    \label{tab:rmse_deltapes}
\end{table}

\begin{figure}[htbp!]
    \includegraphics[width=1.0\columnwidth]{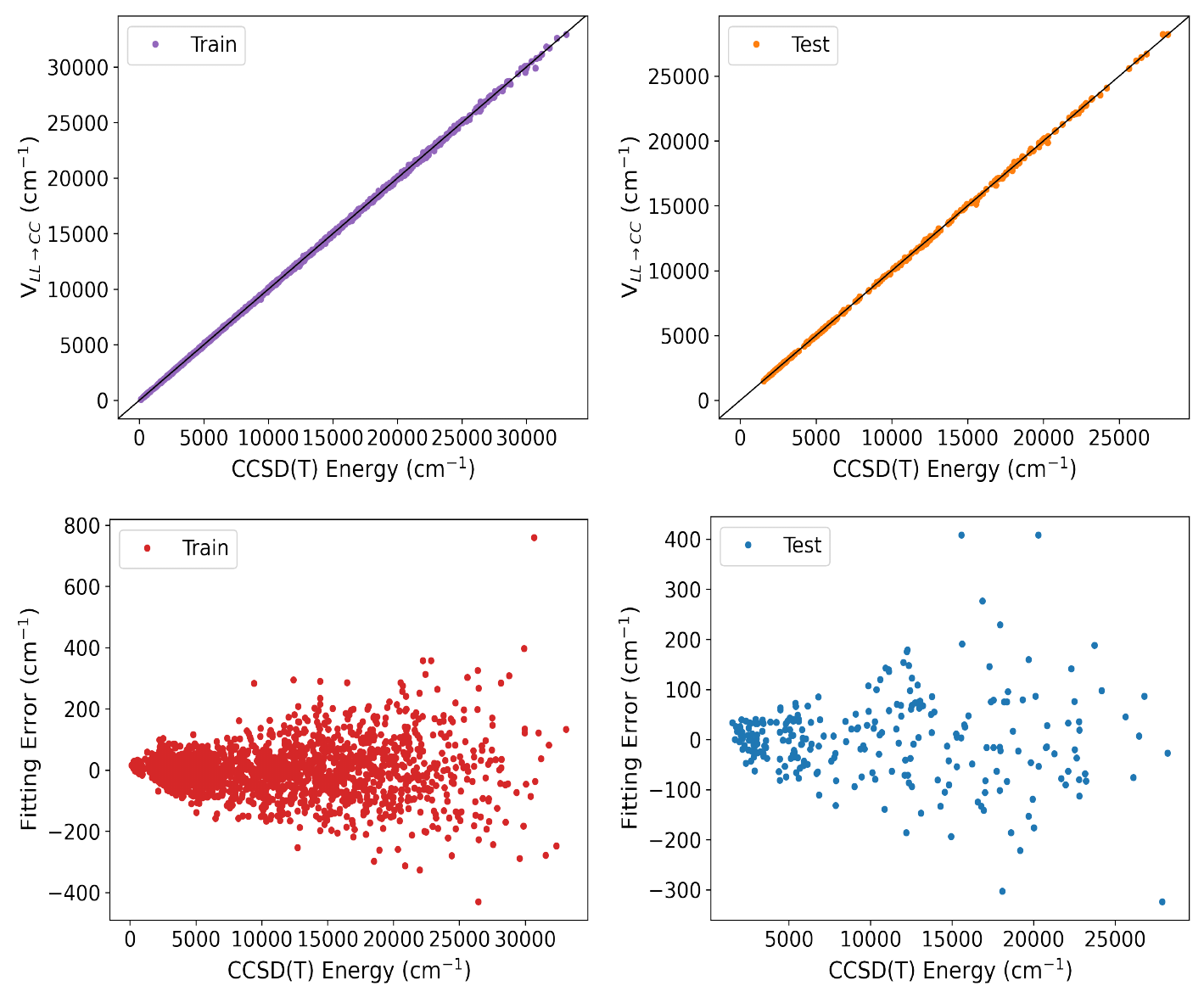}
    \caption{Two upper panels show energies of ethanol from $V_{LL\rightarrow CC}$ vs direct CCSD(T)
ones for the indicated data sets calculated using the PBE functional. Corresponding fitting errors relative to the minimum energy are given in the lower panels.}
        \label{fig:pbe}
\end{figure}

\begin{figure}[htbp!]
    \centering
    \includegraphics[width=1.0\columnwidth]{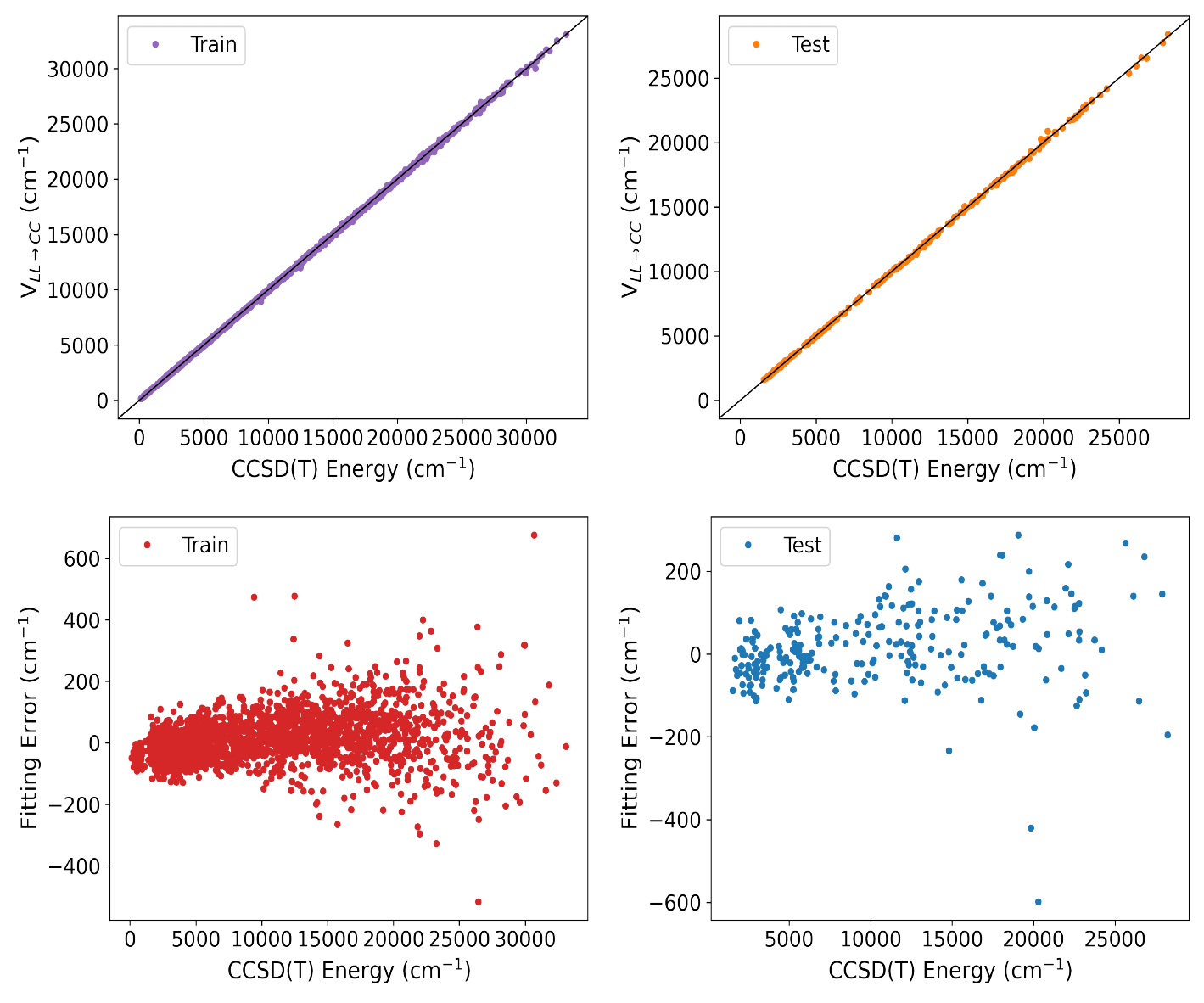}
    \caption{Two upper panels show energies of ethanol from $V_{LL\rightarrow CC}$ vs direct CCSD(T)
ones for the indicated data sets calculated using the M06 functional. Corresponding fitting errors relative to the minimum energy are given in the lower panels.}
    \label{fig:m06}
\end{figure}

\begin{table}[h]
    \centering
    \begin{tabular}{c|c|c|c|c|c}
    \hline \hline
                 & PBE  & M06 & M06-2X & B3LYP & PBE0+MBD \\ [0.5ex] \hline    
    Training     &  78  &  79 &   56  &  53 & 53\\
    Test         &  87  &  97 &   62  &  52 & 52\\ [0.5ex]
        \hline \hline
    \end{tabular}
    \caption{The RMS Fitting Error (in cm$^{-1}$) of corrected PESs $V_{LL \rightarrow CC}$ for training and test data sets.}
    \label{tab:rmse_corrected_pes}
\end{table}
Finally, to obtain the CCSD(T) energies, we incorporate the correction $\Delta V_{CC-LL}$ to the low-level DFT PES, $V_{LL}$. The correlation plots of the $V_{LL \rightarrow CC}$ fit for a training set of 2069 points and a test set of 250 points for the PBE and M06 DFT functional are presented in Figure~\ref{fig:pbe} and ~\ref{fig:m06}, respectively. The RMS training and test errors for the energies of $\Delta$-corrected PES are shown in Table~\ref{tab:rmse_corrected_pes}.

To determine the accuracy of the $V_{LL \rightarrow CC}$ PES for various DFT functionals, we perform the geometry optimization and normal-mode frequency analysis for both \textit{trans} and \textit{gauche} isomers and their two isomerization
saddle point geometries (Anti and Syn). The structures of these isomers and the saddle points are shown in Figure \ref{fig:geom}. The energetics of all four stationary points of ethanol relative to the trans minima, calculated using various DFT functional, are listed in Table~\ref{tab:dftenrg}. The $\Delta$-corrected PES leads to better optimized energetics for all four stationary points across all DFT functionals, as mentioned in Table~\ref{tab:enrgcompare}.

\begin{table}[htbp!]
\centering
\caption{Comparison of the energetics in kcal/mol (cm$^{-1}$) of all four stationary points of ethanol relative to the $trans$ minima from direct DFT calculations.}
\label{tab:dftenrg}
	\begin{tabular*}{1\columnwidth}{@{\extracolsep{\fill}}llllll}
	\hline
	\hline\noalign{\smallskip}
	 Isomer & PBE & M06 &  M06-2X & B3LYP &  PBE0+MBD  \\
 \noalign{\smallskip}
	\noalign{\smallskip}\hline\noalign{\smallskip}
 $Trans$    &  0.00 (0)    & 0.00 (0)   & 0.00 (0)   & 0.00 (0)   &  0.00 (0) \\
 $Gauche$   & -0.37 (-129) & 0.37 (129) & 0.08 (28)  & 0.05 (18)  &  0.05 (19) \\
 TS1 (Anti) & 1.98 (692)   & 1.17 (409) & 1.16 (407) & 1.05 (367) &  1.18 (411) \\
 TS2 (Syn)  & 1.24 (432)   & 1.69 (591) & 1.45 (507) & 1.44 (505) &  1.13 (395) \\

    \noalign{\smallskip}\hline
	\hline
	\end{tabular*}
\end{table}

\begin{table}[htbp!]
\centering
\caption{Comparison of the energetics in kcal/mol (cm$^{-1}$) of all four stationary points of ethanol relative to the $trans$ minima for direct CCSD(T) and $\Delta$-ML PESs.}
\label{tab:enrgcompare}

	\begin{tabular*}{1\columnwidth}{@{\extracolsep{\fill}}lcccccc}
	\hline
	\hline\noalign{\smallskip}

& & \multicolumn{5}{c}{$V_{LL{\rightarrow}CC}$} \\
	\noalign{\smallskip} \cline{3-7} \noalign{\smallskip}
             &  Direct & LL = & LL = & LL = & LL = & LL = \\
	 Isomer & CCSD(T) & PBE & M06 &  M06-2X & B3LYP & PBE0+MBD  \\
 \noalign{\smallskip}
	\noalign{\smallskip}\hline\noalign{\smallskip}
 $Trans$    & 0.00 (0)   &  0.00 (0)  & 0.00 (0)   & 0.00 (0)   & 0.00 (0)   & 0 (0.00) \\
 $Gauche$   & 0.13 (45)  & 0.04 (14)  & 0.21 (73)  & 0.13 (45)  & 0.11 (38) & 0.14 (51) \\
 TS1 (Anti) & 1.09 (381) & 1.22 (427) & 1.12 (392) & 1.08 (378) & 1.08 (378) & 1.04 (363) \\
 TS2 (Syn)  & 1.36 (476) & 1.41 (493) & 1.27 (444) & 1.34 (469) & 1.35 (472) & 1.24 (435) \\

    \noalign{\smallskip}\hline
	\hline
	\end{tabular*}
\end{table}

\begin{figure}
    \centering
    \includegraphics[width = \textwidth]{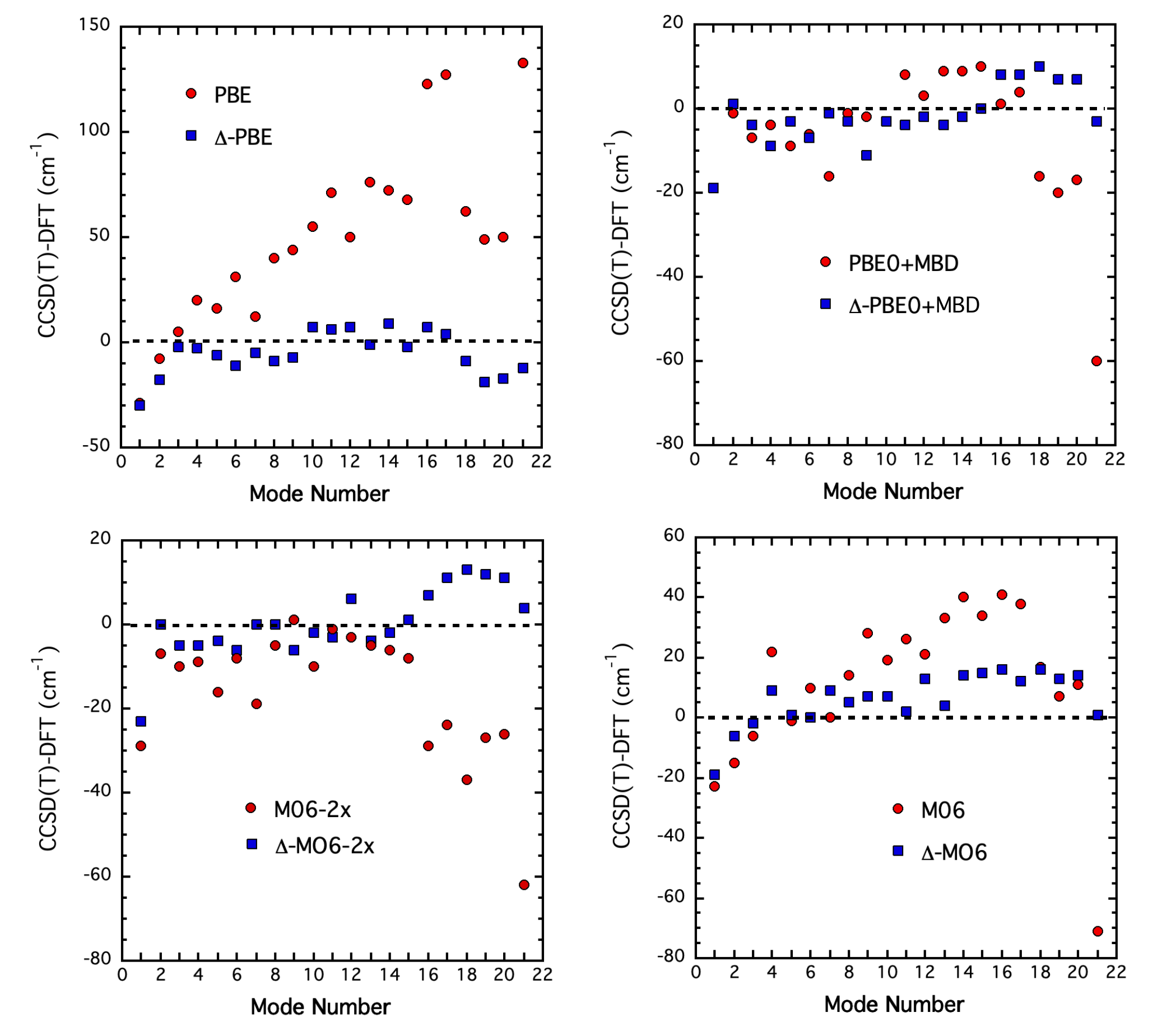}
    \caption{Differences of the CCSD(T) and DFT frequencies (in red) and $\Delta$-corrected frequencies (in blue) for $trans$-ethanol for indicated functionals.}
    \label{fig:freq_plot}
\end{figure}

The comparison of harmonic mode frequencies of various $\Delta$-corrected PES calculated using different DFT PESs (V$_{LL}$) for \textit{trans} ethanol with the corresponding direct CCSD(T) frequencies are shown in Table~\ref{tab:hf_pes}. The overall agreement of these harmonic frequencies with the direct CCSD(T) ones is excellent, as presented in Figure~\ref{fig:freq_plot}. Note that the $\Delta$-corrected PES tends to minimize the gap between the direct-CCSD(T) frequencies and the calculated ones, especially for the high-frequency modes. As depicted in Figure~\ref{fig:freq_plot}, although PBE functional has the highest deviation in frequency from the CCSD(T) values, the correction tends to reduce the deviation within a few cm$^{-1}$.

\begin{table}[htbp!]
\centering
\begin{tabular*}{1.0\columnwidth}{@{\extracolsep{\fill}}rrrrrrrrrrrr}
\hline
\hline\noalign{\smallskip}
& CCSD(T)& \multicolumn{2}{c}{PBE} &  \multicolumn{2}{c}{M06} & \multicolumn{2}{c}{M06-2X}& \multicolumn{2}{c}{B3LYP} & \multicolumn{2}{c}{PBE0+MBD} \\
	\noalign{\smallskip} \cline{2-2} \cline{3-4} \cline{5-6} \cline{7-8} \cline{9-10} \cline{11-12} \noalign{\smallskip}
Mode & Direct & $V_{LL}$ & $\Delta$ML & $V_{LL}$ & $\Delta$ML & $V_{LL}$ & $\Delta$ML & $V_{LL}$ & $\Delta$ML & $V_{LL}$ & $\Delta$ML \\
\hline 
1  & 222  & 251  & 252  & 245  & 241  & 251  & 245  & 237  & 243  & 241  & 241  \\
2  & 274  & 282  & 292  & 289  & 280  & 281  & 274  & 269  & 273  & 275  & 273  \\
3  & 413  & 408  & 415  & 419  & 415  & 423  & 418  & 417  & 417  & 420  & 417  \\
4  & 813  & 793  & 816  & 791  & 804  & 822  & 818  & 820  & 818  & 817  & 822  \\
5  & 907  & 891  & 913  & 908  & 906  & 923  & 911  & 896  & 909  & 916  & 910  \\
6  & 1049 & 1018 & 1060 & 1039 & 1049 & 1057 & 1055 & 1035 & 1055 & 1055 & 1056 \\
7  & 1115 & 1103 & 1120 & 1115 & 1106 & 1134 & 1115 & 1094 & 1115 & 1131 & 1116 \\
8  & 1180 & 1140 & 1189 & 1166 & 1175 & 1185 & 1180	& 1176 & 1181 & 1181 & 1183 \\
9  & 1274 & 1230 & 1281 & 1246 & 1267 & 1273 & 1280 & 1266 & 1284 & 1276 & 1285 \\
10 & 1300 & 1245 & 1293 & 1281 & 1293 & 1310 & 1302 & 1299 & 1302 & 1303 & 1303 \\
11 & 1402 & 1331 & 1396 & 1376 & 1400 & 1403 & 1405 & 1402 & 1403 & 1394 & 1406 \\
12 & 1456 & 1406 & 1449 & 1435 & 1443 & 1459 & 1450 & 1446 & 1454 & 1453 & 1458 \\
13 & 1484 & 1408 & 1485 & 1451 & 1480 & 1489 & 1488 & 1483 & 1488 & 1475 & 1488 \\
14 & 1501 & 1429 & 1492 & 1461 & 1487 & 1507 & 1503 & 1498 & 1500 & 1492 & 1503 \\
15 & 1531 & 1463 & 1533 & 1497 & 1516 & 1539 & 1530 & 1524 & 1530 & 1521 & 1531 \\
16 & 3001 & 2878 & 2994 & 2960 & 2985 & 3030 & 2994 & 2978 & 2995 & 3000 & 2993 \\
17 & 3036 & 2909 & 3032 & 2998 & 3024 & 3060 & 3025 & 3005 & 3028 & 3032 & 3028 \\
18 & 3042 & 2980 & 3051 & 3025 & 3026 & 3079 & 3029 & 3031 & 3036 & 3058 & 3032 \\
19 & 3122 & 3073 & 3141 & 3115 & 3109 & 3149 & 3110 & 3098 & 3120 & 3142 & 3115 \\
20 & 3127 & 3077 & 3144 & 3116 & 3113 & 3153 & 3116 & 3105 & 3126 & 3144 & 3120 \\
21 & 3853 & 3720 & 3865 & 3924 & 3852 & 3915 & 3849 & 3843 & 3862 & 3913 & 3856 \\
\hline 
MAE &   & 54    & 9 & 23 & 9 &16 & 6 & 11 & 4 & 11 & 6\\
\hline
\end{tabular*}
\caption{Comparison of Harmonic Frequencies (in cm$^{-1}$) of DFT PESs and $\Delta$-corrected PES computed using indicated DFT functionals and corresponding \textit{ab initio} ones (CCSD(T)-F12a/aug-cc-pVDZ) for \textit{trans}-ethanol.}
\label{tab:hf_pes}
\end{table}

\begin{figure}[htbp!]
    \centering
    \includegraphics[width =0.95 \textwidth]{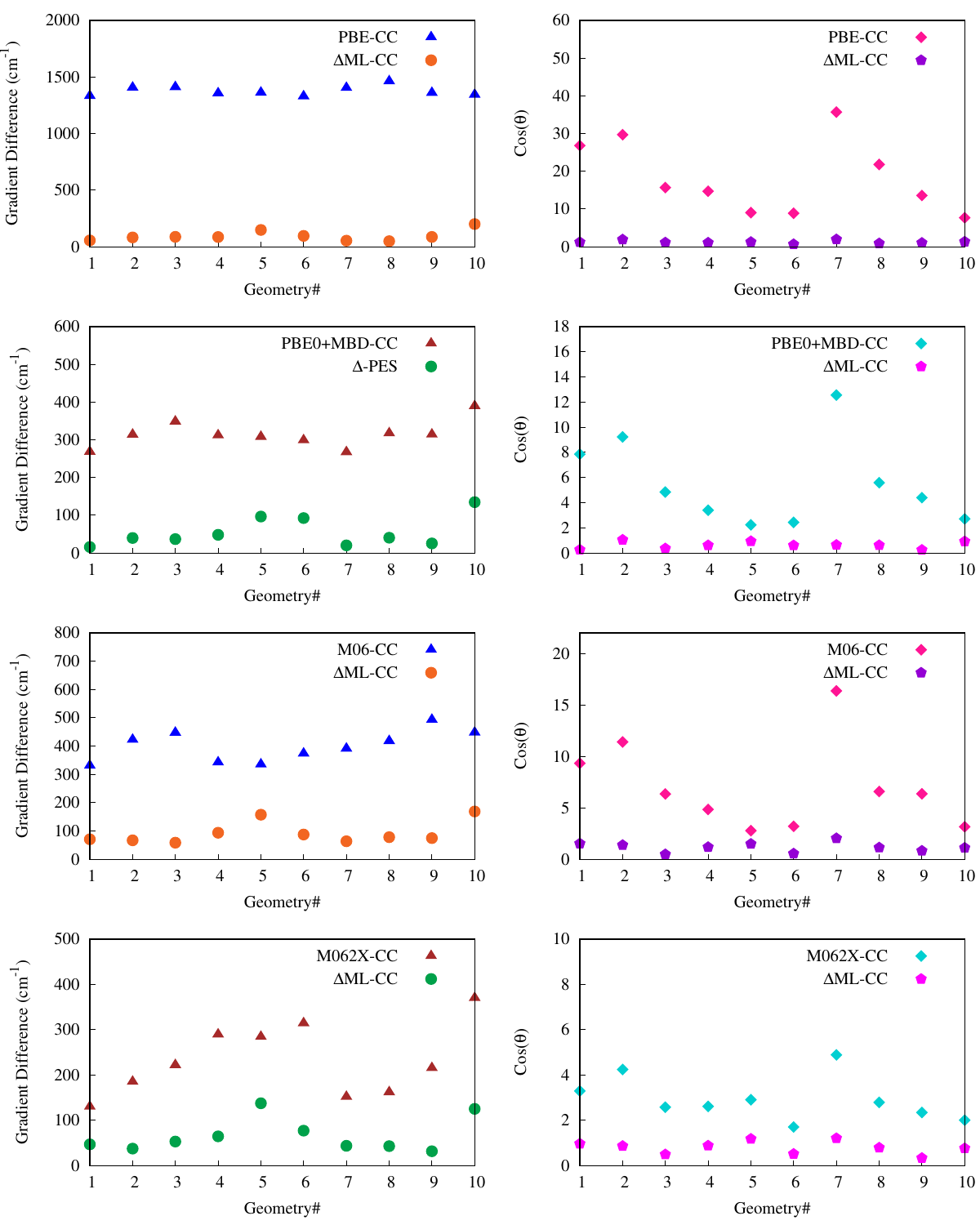}
    \caption{Plot of mean absolute gradient magnitude difference (left panel) and $\cos\theta$ (right panel), where $\theta$ is angle between the direct CCSD(T) and indicated DFT gradients as well as corresponding $V_{LL{\rightarrow}CC}$ PES gradients for randomly selected 10 ethanol geometries. See the text for details.}
    \label{fig:grad_plot}
\end{figure}

Furthermore, we examine the change in the PES gradient after the incorporation of the correction. In order to make a more detailed examination of the errors in gradients, we calculated cosine of the angle between the direct CCSD(T) and DFT gradient vector as well as corresponding $V_{LL{\rightarrow}CC}$ PES gradient vector, also the mean absolute difference (magnitude of 27 gradient components for each geometry) between these two gradient vectors for 10 randomly selected geometries. This is shown in Figure~\ref{fig:grad_plot}.
As seen, there is a substantial reduction in the errors in the gradient in the $\Delta$-ML corrected PES compared to the DFT PESs. Specially, in case of PBE the the gradient differences are much larger as well as the cos$\theta$ values.
This is especially encouraging as the correction PES, $\Delta{V_{CC-LL}}$, is trained only on CCSD(T) energies without CCSD(T) gradients. Presumably, including gradient data in the training of $\Delta{V_{CC-LL}}$ would result in a larger reduction in the error. We plan to investigate this in the future.

\begin{figure}
    \centering
    \includegraphics{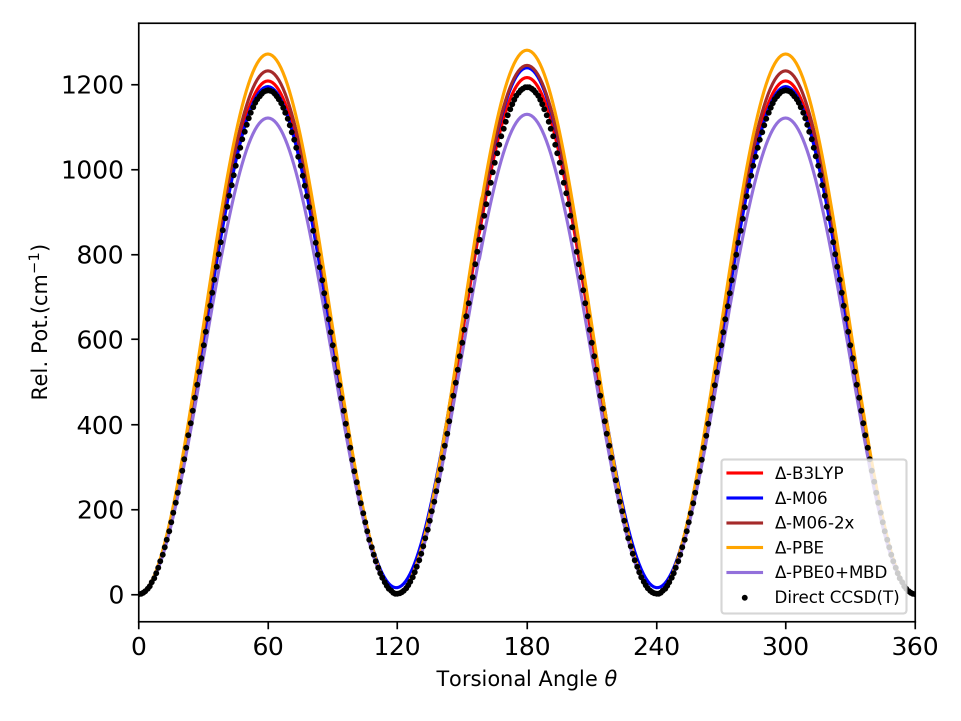}
    \caption{Comparison of torsional potential (not fully relaxed) of the methyl rotor of trans ethanol between direct CCSD(T) and $\Delta$-corrected PES computed using indicated DFT functionals.}
    \label{fig:rotor_energy}
\end{figure}

Moreover, we compare the PES calculated torsional barrier for the methyl rotor with the direct CCSD(T) level. The methyl rotor torsional potentials (not fully relaxed) for the \textit{trans} minima as a function of the torsional angle are shown in Figure~\ref{fig:rotor_energy}. For all the DFT functionals, the results from the $\Delta$-corrected PESs are comparable to the direct \textit{ab initio} calculations at the CCSD(T) level, as mentioned in Table~\ref{tab:torsional_methyl}. Note that the methyl torsional barrier height for the \textit{trans} isomer evaluated from the microwave spectroscopy is 1174 cm$^{-1}$~\cite{Quade2000,Pearson1995}. Similarly, another experimental analysis of the infrared and Raman spectra determined the methyl torsional barriers to be 1185 cm$^{-1}$~\cite{Durig1990}.

\begin{table}[htbp!]
    \centering
    \begin{tabular}{c|c|c|c|c|c} \hline \hline
 Direct-CCSD(T) & $\Delta$-PBE & $\Delta$-M06 & $\Delta$-M06-2x & $\Delta$-B3LYP & $\Delta$-PBE0+MBD  \\  [0.5ex] \hline    
         1194 & 1272 & 1195 & 1232 & 1208 & 1121\\
         \hline \hline
    \end{tabular}
    \caption{Barrier height of the methyl rotor torsional potential for the trans isomer. Energies are in cm$^{-1}$. }
    \label{tab:torsional_methyl}
\end{table}

\subsection{$\Delta$-ML for Force Field}
\begin{figure}
    \centering
    \includegraphics[width=1.0\textwidth]{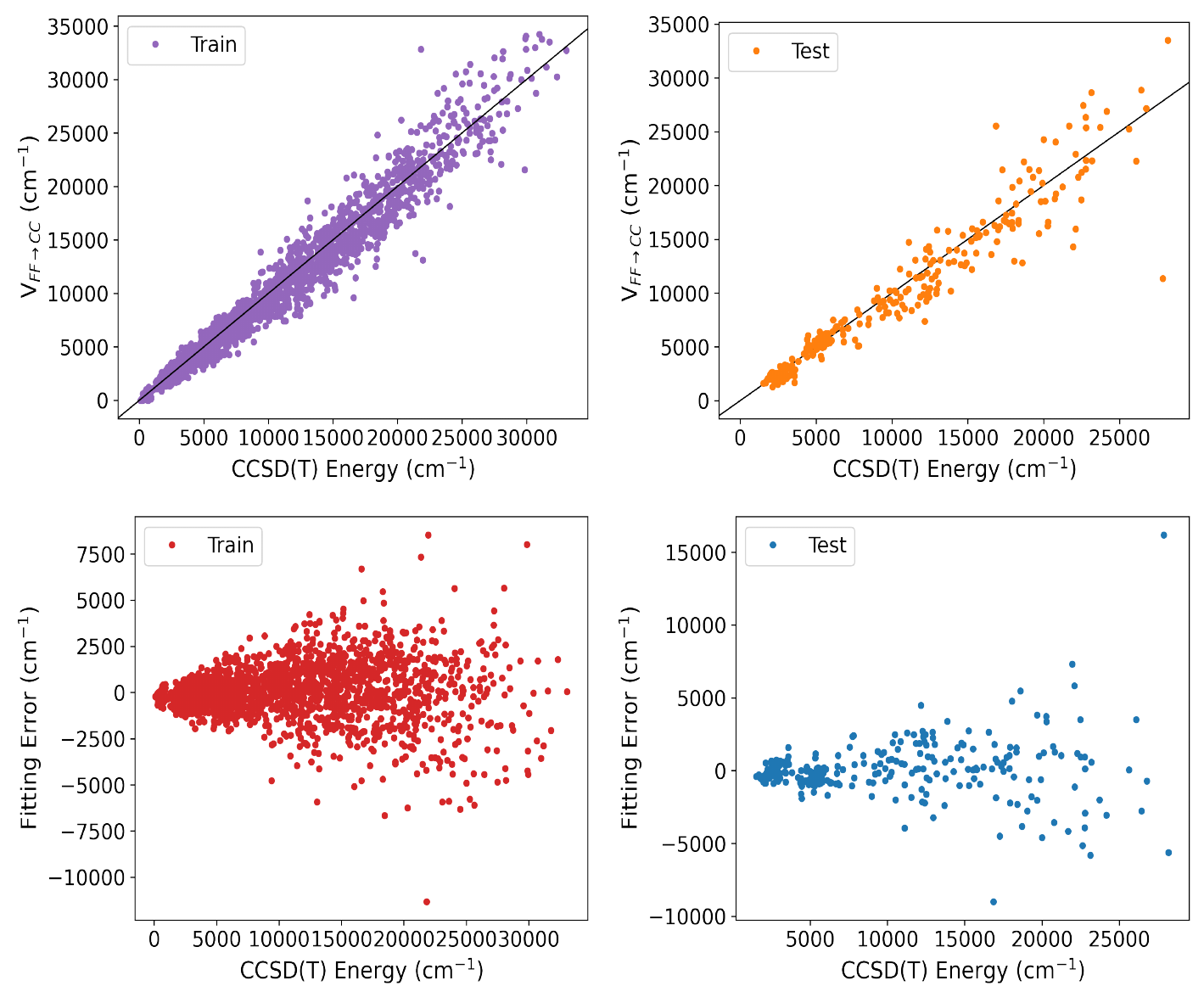}
    \caption{Two upper panels show energies of ethanol from V$_{FF \rightarrow CC}$ vs direct CCSD(T) ones for the indicated data sets calculated using the Harmonic approximation for the MP2 corrected force field. Corresponding fitting errors relative to the minimum energy are given in Table \ref{tab:ff_rmse}.}
    \label{fig:ff_ho}
\end{figure}

To calculate the $\Delta$-corrected force field potential, we first calculated the force field potential energy using Eqn.~\ref{eqn:ff_potn}.
Next, we train the $\Delta$-correction PES on the difference between the CCSD(T) and FF energies of 2069 geometries and test the obtained surface on the remaining 250 geometries.\cite{nandi2022quantum} To fit the corrected PES, a maximum polynomial order of 2  is used with permutationally symmetry 321111 for the training data set.
A plot of V$_{FF \rightarrow CC}$ versus corresponding direct CCSD(T) energies for the training and test sets calculated using the harmonic approximation for the MP2 corrected force field, along with the fitting error, is shown in Figure~\ref{fig:ff_ho}. A huge fitting error for both the training and test sets is found, with RMSE values of 1436 cm$^{-1}$ and 2097 cm$^{-1}$, respectively. The substantial RMSE observed in the $\Delta$-ML corrected force field PES indicates the imprecise fitting. 

The harmonic approximation works well for the small oscillation around the equilibrium position but its accuracy decreases for larger amplitude vibrations where anharmonicity becomes significant. Hence, we use a Morse potential as $V_{bond}$ to provide a more realistic representation to the higher bond stretching.

\begin{align}
    V_{bond} = D_e(1-e^{-\alpha(r-r_e)})^2
    \label{eqn:morse}
\end{align}

Here, the approximate value of $\alpha$ is equal to $\sqrt{{k_{bond}}/{D_e}}$ for all bond types, with the dissociation energy $D_e$ provided in Table~\ref{tab:ff_parameters}.

\begin{table}[htbp!]
    \centering
    \begin{tabular}{c|c|c|c} \hline
       Bond type  & $r_e$ ($\AA$) & $k_{bond}$ ($kcal/mol \AA^2$) & $D_e$ ($kcal/mol$)\\
       \hline
        \ce{C-C} & 1.5204 & 551.9110  & 82.69\\
        \ce{O-H} & 0.9609 & 1056.6764 & 110.66\\
        \ce{C-O} & 1.4396 & 577.2346  & 85.56\\
        \ce{C-H} & 1.0937 & 742.5561  & 98.71\\ 
        \hline
    \end{tabular}
    \caption{Intramolecular potential parameters of ethanol taken from ref.~\citenum{rogers2020accurate}.}
    \label{tab:ff_parameters}
\end{table}

\begin{figure}
    \centering
    \includegraphics[width=1.0\textwidth]{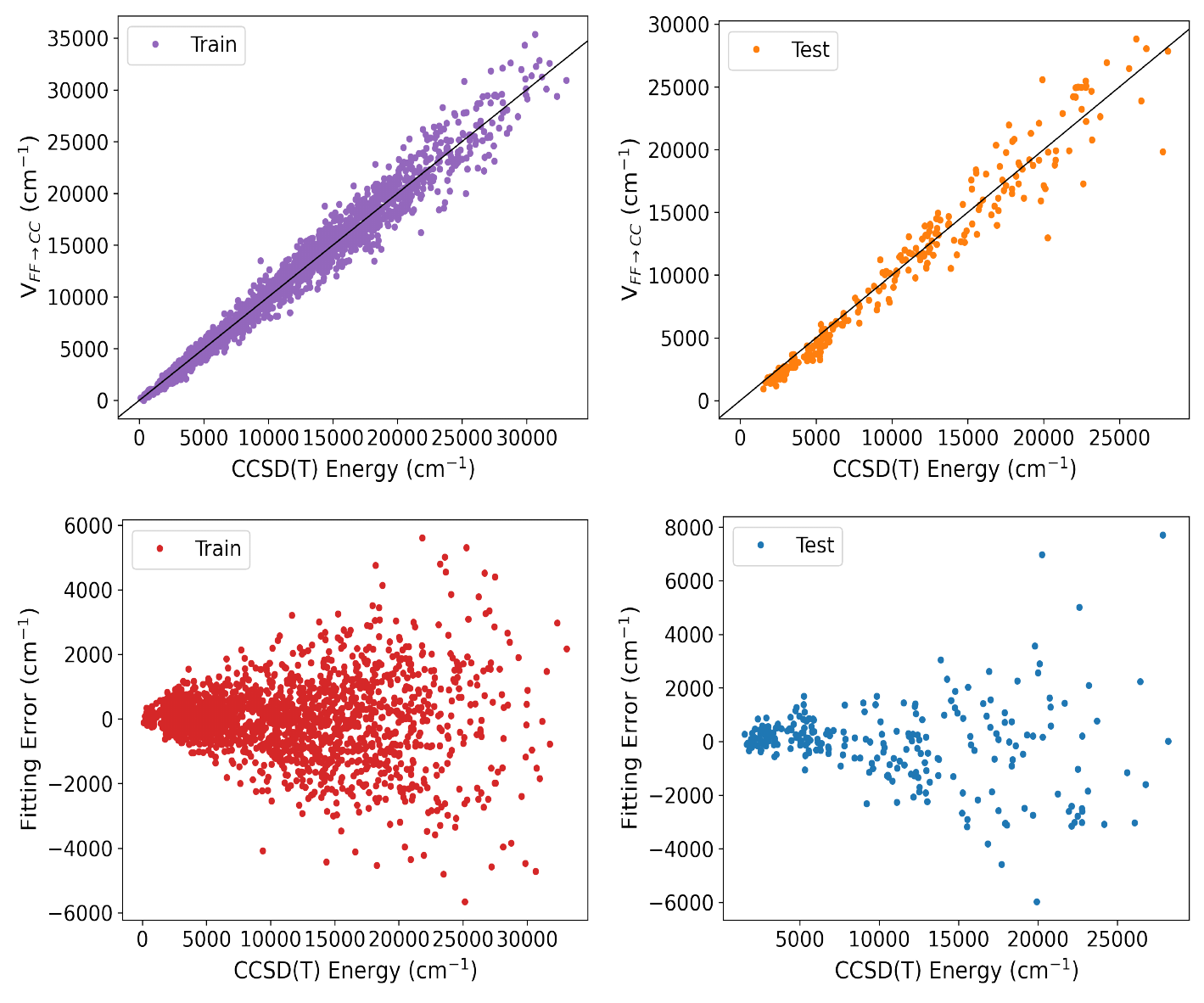}
    \caption{Two upper panels show energies of ethanol from V$_{FF \rightarrow CC}$ vs direct CCSD(T) ones for the indicated data sets calculated using the Morse potential for the MP2 corrected force field. Corresponding fitting errors relative to the minimum energy are given in Table \ref{tab:ff_rmse}.}
    \label{fig:ff_mo}
\end{figure}

A plot of V$_{FF \rightarrow CC}$ versus corresponding direct CCSD(T) energies for the training and test sets, calculated using the Morse potential along with the fitting error, is shown in Figure~\ref{fig:ff_mo}.  The fitting error decreases slightly for both the training and test sets compared to Figure~\ref{fig:ff_ho}, with reduced RMSE values of 1089 cm$^{-1}$ and 1529 cm$^{-1}$, respectively. However, these RMSE values are still large enough to produce inaccurate results for the entire dataset. Therefore, we attempt to improve the fitting by implementing energy cut-offs across the entire dataset. For this purpose, we select two energy cut-offs at 10000 cm$^{-1}$ and 5000 cm$^{-1}$ above the global minimum. For the 10000 cm$^{-1}$ energy cut off case, the correction PES is trained on the difference between the CCSD(T) and FF absolute energies of 1124 geometries and tested on the remaining 125 geometries. For the 5000 cm$^{-1}$ energy cut off case, the training and test geometries are 702 and 65, respectively.

Table~\ref{tab:ff_rmse} presents the RMS errors in $\Delta$-corrected PES computed using the force field for both the training and test datasets. The RMSE values decrease by a factor of 5 for the energy cut-off set at 10000 cm$^{-1}$ and by a factor of 15 for the dataset at 5000 cm$^{-1}$, compared to the RMSE of the $\Delta$-corrected PES computed using the force field with the harmonic approximation. Since direct CCSD(T) energies of the ethanol isomers and their saddle point transition states are quite small in comparison to the RMSE values, their energy optimization results are random. Note that the time taken to calculate 100,000 data points using the force field is 2.04 and 2.09 sec for the harmonic and Morse potentials, respectively. Even after adding the $\Delta$-correction, the time taken to calculate 100,000 data points is 2.15 sec. Hence, the force field $\Delta$-ML PES is much faster than the one using the DFT functional; to be precise, it has almost doubled the evaluation speed.

\begin{table}[ht]
    \centering
    \begin{tabular}{l|c|c|c|c}
         RMS Error  & Harmonic FF$^a$ & Morse FF$^a$ & Morse FF$^b$ & Morse FF$^c$  \\ \hline
         Train & 1436 & 1089 & 294 & 98\\
         Test  & 2097 & 1529 & 462 & 233\\
         \hline
    \end{tabular}
      \\
         $^a$ Fit using all CCSD(T) data to roughly 32000 cm$^{-1}$   \\
         $^b$ Fit using data at 10000 cm$^{-1}$  \\
         $^c$ Fit using data at 5000 cm$^{-1}$\\
    \caption{The RMS error in $\Delta$-corrected energies (in cm$^{-1}$) computed using the force field with the original harmonic stretch and present Morse potential modified stretch potentials for train and test data sets. These calculations use the fitting basis of 208 terms described in the text. }
    \label{tab:ff_rmse}
\end{table}

Next, we performed normal-mode analyses for $trans$-ethanol to examine the vibrational frequency predictions of these PESs. The comparisons of harmonic frequencies with the corresponding \textit{ab initio} frequencies for the $trans$-ethanol are shown in Table~\ref{tab:ff_freq}.  As seen in the table, the original FF produces poor results except for the two lowest frequencies.  The sets of corrected FFs all show significant improvement at these frequencies. The corrections to the original FF, denoted as the ``Harmonic FF", and those where the harmonic stretch modes were replaced by Morse potentials, denoted as the ``Morse FF", were evaluated.  Overall, the corrected Morse FF results are superior to those of the corrected harmonic FF. Notably, there are interesting dependencies to the extent of the training dataset.  Limiting the maximum energy to 5000 cm$^{-1}$  produces the best correction, and this is for the Morse FF. This is probably due to the higher precision for the correction PES for this limited energy range, as shown in Table \ref{tab:ff_rmse}. However, the results using the full range still show a significant improvement over the uncorrected FF, with the mean absolute error (MAE) being approximately five times less than that of the uncorrected FF.  The reason for this can be deduced from Figures. \ref{fig:ff_ho} and \ref{fig:ff_mo}.  As seen, the fitting errors are relatively small for energies up to 10 000 cm$^{-1}$ and then grow rapidly above that energy. Therefore, for properties that are largely determined by energies up to 10 000 cm$^{-1}$, such as harmonic frequencies and torsional barriers the correction PES trained on this energy range performs well.
\begin{table}[htbp!]
\centering
\begin{tabular*}{1.0\columnwidth}{@{\extracolsep{\fill}}rrrrrrrrr}
\hline
\hline\noalign{\smallskip}
& CCSD(T)& Force Field & {Harmonic FF} &  \multicolumn{3}{c}{Morse FF}  \\
	\noalign{\smallskip} \cline{2-2}  \cline{3-3}\cline{4-4} \cline{5-7}  \noalign{\smallskip}
Mode & Direct & Harmonic & $\Delta$ML$^a$ & $\Delta$ML$^a$ & $\Delta$ML$^b$ & $\Delta$ML$^c$ \\
\hline 
1  & 222  & 234& 258&   381 & 245 & 234  \\
2  & 274  & 261& 276&   385 & 277 & 240\\
3  & 413  & 606& 400&   507 & 351 & 402\\
4  & 813  & 1131& 631&  747 & 748 & 779\\
5  & 907  & 1148& 829&  875 & 917 & 960\\
6  & 1049 & 1300& 835&  992 & 1087& 1083\\
7  & 1115 & 1421& 1135& 1055& 1148& 1144 \\
8  & 1180 & 1443& 1320& 1090& 1165& 1191\\
9  & 1274 & 1753& 1322& 1158& 1224& 1298 \\
10 & 1300 & 1816& 1453& 1190 & 1314& 1325\\
11 & 1402 & 1960& 1481& 1195 & 1347& 1413\\
12 & 1456 & 1994& 1564& 1266 & 1390& 1454\\
13 & 1484 & 2019& 1585& 1421 & 1423& 1530\\
14 & 1501 & 2049& 1600& 1480 & 1427& 1539\\
15 & 1531 & 2142& 1786& 1484 & 1508& 1657\\
16 & 3001 & 4247& 1933& 3353 & 3035& 3112\\
17 & 3036 & 4300& 2200& 3442 & 3224& 3135\\
18 & 3042 & 4398& 2256& 3468 & 3295& 3173\\
19 & 3122 & 4409& 2420& 3476 & 3316& 3263\\
20 & 3127 & 4410& 2674& 3502 & 3322& 3304\\
21 & 3853 & 5129& 3524& 4099 & 4145& 3930\\
\hline 
MAE &     &624 & 272 & 116 & 111 & 171 \\
\hline
\end{tabular*}
\\
         $^a$ Fit using full data points up to 35000 cm$^{-1}$  \\
         $^b$ Fit using data points up to 10000 cm$^{-1}$\\
         $^c$ Fit using data points up to 5000 cm$^{-1}$\\
\caption{Comparison of Harmonic Frequencies (in cm$^{-1}$) between V$_{FF \rightarrow CC}$ PES computed at indicated force field and corresponding \textit{ab initio} ones (CCSD(T)-F12a/aug-cc-pVDZ) for $trans$-ethanol.}
\label{tab:ff_freq}
\end{table}

\begin{figure}
    \centering
    \includegraphics{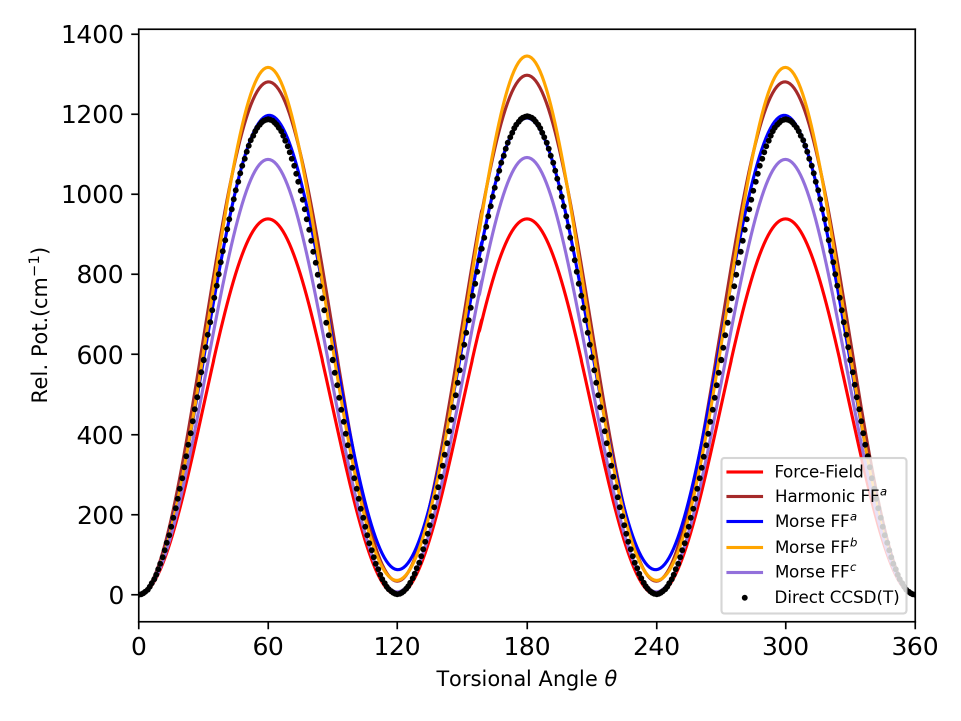}
    \caption{Comparison of torsional potential (not fully relaxed) of the methyl rotor of trans-ethanol between direct CCSD(T) and $\Delta$-corrected  PES computed using the force field.}
    \label{fig:rotor_energy_ff}
\end{figure}

Lastly, we analyzed the torsional barrier for the methyl rotor calculated by $\Delta$-ML PES using the force field. The results of the methyl torsional barrier height for the trans isomer, calculated from the various PESs of the force field, are listed in Table~\ref{tab:torsional_methyl_ff}. As shown in Figure~\ref{fig:rotor_energy_ff}, the torsional barrier height for the harmonic force field is much lower than the direct CCSD(T) value. For all the corrected PESs, the barrier height improves. The barrier height matches to the direct CCSD(T) value for the Morse FF when the full dataset is considered.

\begin{table}[htbp!]
\centering
\begin{tabular*}{1.0\columnwidth}{@{\extracolsep{\fill}}cccccccc}
\hline
\hline\noalign{\smallskip}
 CCSD(T)& Force Field & {Harmonic FF} &  \multicolumn{3}{c}{Morse FF}  \\
	\noalign{\smallskip} \cline{1-1}  \cline{2-2}\cline{3-3} \cline{4-6}  \noalign{\smallskip}
 Direct & Harmonic & $\Delta$ML$^a$  & $\Delta$ML$^a$ & $\Delta$ML$^b$ & $\Delta$ML$^c$ \\
\hline 
 1194 & 938 & 1280 & 1194 & 1316 & 1086 \\
\hline 
\end{tabular*}
\\
         $^a$ Fit using full data points up to 35000 cm$^{-1}$  \\
         $^b$ Fit using data points up to 10000 cm$^{-1}$\\
         $^c$ Fit using data points up to 5000 cm$^{-1}$\\
    \caption{Barrier height of the methyl rotor torsional potential calculated using $\Delta$-corrected force field for the trans isomer. Energies are in cm$^{-1}$. }
    \label{tab:torsional_methyl_ff}
\end{table}

Overall, the correction to this classical FF has been successful.  And, it is reasonable to ask how the approach taken could be used for general classical FFs, especially for molecules much larger than ethanol. There isn't a simple answer to this question, but clearly this is a fruitful area for future work.  One preliminary thought is to make use of the simple form of FFs, which is just the generalization of Equations (4)-(6), and to correct groups of terms instead of the entire FF. Such an approach would be ``universal" and transferable.

\newpage

\section{Summary and Conclusions}

The generality of the single-step $\Delta$-ML method we proposed and applied using B3LYP to a number of PIP PESs has been demonstrated here for ethanol using other popular DFT functionals. In each case, the $\Delta$-ML method produces a substantial improvement in accuracy compared to the CCSD(T) benchmark results. The most dramatic improvement is observed in the harmonic frequencies, where the DFT PIP PESs produce both significant underestimates and overestimates of the CH and OH-stretch frequencies. Additionally, we achieved significant improvement over DFT gradients without using CC gradients data to correct the PES.
An exploratory application of this $\Delta$-ML method to a recent force field (FF) for ethanol was given.  Notably, the inaccurate harmonic frequencies at the global minimum from the force field are significantly corrected.  The torsional barrier from the FF is also improved using the $\Delta$-ML method.  Additionally, the computational cost for the correction is about the same as the cost to evaluate the simple FF.  

The new DFT $\Delta$-ML potentials are expected to perform as well for Diffusion Monte Carlo and VSCF/VCI calculations as the original B3LYP $\Delta$-ML one.\cite{nandi2022quantum,ethanolMM}  However, the performance of the $\Delta$-ML corrected Force Field will need to be investigated for such calculations. In addition, it will also be of interest to test the new  Quantum-Monte-Carlo based sGDML potential\cite{slootman2024accurate} for such calculations.

Finally, this one-step $\Delta$-ML method is very straightforward and can be easily implemented into other ML methods or descriptors. While ethanol molecule is used here as a prototype example, this approach is also applicable to large molecular systems for developing machine-learned force fields as accurate as the CC level.

\begin{acknowledgement}
AN and AT acknowledge support from PHANTASTIC grant INTER/MERA22/16521502/PHANTASTIC.
JMB and PP acknowledge support from NASA grant 80NSSC22K1167. RC acknowledges support from Universit\`a degli Studi di Milano under grant PSR2022\_DIP\_005\_PI\_RCONT.
\end{acknowledgement}

\bibliography{refs}

\providecommand{\latin}[1]{#1}
\makeatletter
\providecommand{\doi}
  {\begingroup\let\do\@makeother\dospecials
  \catcode`\{=1 \catcode`\}=2 \doi@aux}
\providecommand{\doi@aux}[1]{\endgroup\texttt{#1}}
\makeatother
\providecommand*\mcitethebibliography{\thebibliography}
\csname @ifundefined\endcsname{endmcitethebibliography}  {\let\endmcitethebibliography\endthebibliography}{}
\begin{mcitethebibliography}{67}
\providecommand*\natexlab[1]{#1}
\providecommand*\mciteSetBstSublistMode[1]{}
\providecommand*\mciteSetBstMaxWidthForm[2]{}
\providecommand*\mciteBstWouldAddEndPuncttrue
  {\def\EndOfBibitem{\unskip.}}
\providecommand*\mciteBstWouldAddEndPunctfalse
  {\let\EndOfBibitem\relax}
\providecommand*\mciteSetBstMidEndSepPunct[3]{}
\providecommand*\mciteSetBstSublistLabelBeginEnd[3]{}
\providecommand*\EndOfBibitem{}
\mciteSetBstSublistMode{f}
\mciteSetBstMaxWidthForm{subitem}{(\alph{mcitesubitemcount})}
\mciteSetBstSublistLabelBeginEnd
  {\mcitemaxwidthsubitemform\space}
  {\relax}
  {\relax}

\bibitem[Bowman \latin{et~al.}(2011)Bowman, Czak{\'o}, and Fu]{bowman11}
Bowman,~J.~M.; Czak{\'o},~G.; Fu,~B. High-dimensional ab initio potential energy surfaces for reaction dynamics calculations. \emph{Phys. Chem. Chem. Phys.} \textbf{2011}, \emph{13}, 8094--8111\relax
\mciteBstWouldAddEndPuncttrue
\mciteSetBstMidEndSepPunct{\mcitedefaultmidpunct}
{\mcitedefaultendpunct}{\mcitedefaultseppunct}\relax
\EndOfBibitem
\bibitem[Qu \latin{et~al.}(2018)Qu, Yu, and Bowman]{ARPC2018}
Qu,~C.; Yu,~Q.; Bowman,~J.~M. Permutationally invariant potential energy surfaces. \emph{Annu. Rev. Phys. Chem.} \textbf{2018}, \emph{69}, 6.1--6.25\relax
\mciteBstWouldAddEndPuncttrue
\mciteSetBstMidEndSepPunct{\mcitedefaultmidpunct}
{\mcitedefaultendpunct}{\mcitedefaultseppunct}\relax
\EndOfBibitem
\bibitem[Fu and Zhang(2018)Fu, and Zhang]{Fu18}
Fu,~B.; Zhang,~D.~H. Ab initio potential energy surfaces and quantum dynamics for polyatomic bimolecular reactions. \emph{J. Chem. Theory Comput.} \textbf{2018}, \emph{14}, 2289--2303\relax
\mciteBstWouldAddEndPuncttrue
\mciteSetBstMidEndSepPunct{\mcitedefaultmidpunct}
{\mcitedefaultendpunct}{\mcitedefaultseppunct}\relax
\EndOfBibitem
\bibitem[Jiang \latin{et~al.}(2020)Jiang, Li, and Guo]{guo20}
Jiang,~B.; Li,~J.; Guo,~H. High-Fidelity Potential Energy Surfaces for Gas-Phase and Gas-Surface Scattering Processes from Machine Learning. \emph{J. Phys. Chem. Lett.} \textbf{2020}, \emph{11}, 5120--5131\relax
\mciteBstWouldAddEndPuncttrue
\mciteSetBstMidEndSepPunct{\mcitedefaultmidpunct}
{\mcitedefaultendpunct}{\mcitedefaultseppunct}\relax
\EndOfBibitem
\bibitem[Győri and Czakó(2020)Győri, and Czakó]{robo20}
Győri,~T.; Czakó,~G. Automating the Development of High-Dimensional Reactive Potential Energy Surfaces with the robosurfer Program System. \emph{J. Chem. Theory Comput.} \textbf{2020}, \emph{16}, 51--66\relax
\mciteBstWouldAddEndPuncttrue
\mciteSetBstMidEndSepPunct{\mcitedefaultmidpunct}
{\mcitedefaultendpunct}{\mcitedefaultseppunct}\relax
\EndOfBibitem
\bibitem[Chmiela \latin{et~al.}(2018)Chmiela, Sauceda, M{\"u}ller, and Tkatchenko]{Tkatch2018}
Chmiela,~S.; Sauceda,~H.~E.; M{\"u}ller,~K.-R.; Tkatchenko,~A. Towards exact molecular dynamics simulations with machine-learned force fields. \emph{Nat. Commun.} \textbf{2018}, \emph{9}, 3887\relax
\mciteBstWouldAddEndPuncttrue
\mciteSetBstMidEndSepPunct{\mcitedefaultmidpunct}
{\mcitedefaultendpunct}{\mcitedefaultseppunct}\relax
\EndOfBibitem
\bibitem[Sauceda \latin{et~al.}(2019)Sauceda, Chmiela, Poltavsky, Müller, and Tkatchenko]{Tkatch19}
Sauceda,~H.~E.; Chmiela,~S.; Poltavsky,~I.; Müller,~K.-R.; Tkatchenko,~A. Molecular force fields with gradient-domain machine learning: Construction and application to dynamics of small molecules with coupled cluster forces. \emph{J. Chem. Phys.} \textbf{2019}, \emph{150}, 114102\relax
\mciteBstWouldAddEndPuncttrue
\mciteSetBstMidEndSepPunct{\mcitedefaultmidpunct}
{\mcitedefaultendpunct}{\mcitedefaultseppunct}\relax
\EndOfBibitem
\bibitem[Houston \latin{et~al.}(2022)Houston, Qu, Nandi, Conte, Yu, and Bowman]{Bowman_reverse2022}
Houston,~P.~L.; Qu,~C.; Nandi,~A.; Conte,~R.; Yu,~Q.; Bowman,~J.~M. Permutationally invariant polynomial regression for energies and gradients, using reverse differentiation, achieves orders of magnitude speed-up with high precision compared to other machine learning methods. \emph{J. Chem. Phys.} \textbf{2022}, \emph{156}, 044120\relax
\mciteBstWouldAddEndPuncttrue
\mciteSetBstMidEndSepPunct{\mcitedefaultmidpunct}
{\mcitedefaultendpunct}{\mcitedefaultseppunct}\relax
\EndOfBibitem
\bibitem[Bartók and Csányi(2015)Bartók, and Csányi]{GP-2015-1}
Bartók,~A.~P.; Csányi,~G. Gaussian approximation potentials: A brief tutorial introduction. \emph{Int. J. Quantum Chem.} \textbf{2015}, \emph{115}, 1051--1057\relax
\mciteBstWouldAddEndPuncttrue
\mciteSetBstMidEndSepPunct{\mcitedefaultmidpunct}
{\mcitedefaultendpunct}{\mcitedefaultseppunct}\relax
\EndOfBibitem
\bibitem[Smith \latin{et~al.}(2017)Smith, Isayev, and Roitberg]{AN1}
Smith,~J.~S.; Isayev,~O.; Roitberg,~A.~E. ANI-1: an extensible neural network potential with DFT accuracy at force field computational cost. \emph{Chem. Sci.} \textbf{2017}, \emph{8}, 3192--3203\relax
\mciteBstWouldAddEndPuncttrue
\mciteSetBstMidEndSepPunct{\mcitedefaultmidpunct}
{\mcitedefaultendpunct}{\mcitedefaultseppunct}\relax
\EndOfBibitem
\bibitem[Zhang \latin{et~al.}(2018)Zhang, Han, Wang, Car, and E]{dpmd2018}
Zhang,~L.; Han,~J.; Wang,~H.; Car,~R.; E,~W. Deep Potential Molecular Dynamics: A Scalable Model with the Accuracy of Quantum Mechanics. \emph{Phys. Rev. Lett.} \textbf{2018}, \emph{120}, 143001\relax
\mciteBstWouldAddEndPuncttrue
\mciteSetBstMidEndSepPunct{\mcitedefaultmidpunct}
{\mcitedefaultendpunct}{\mcitedefaultseppunct}\relax
\EndOfBibitem
\bibitem[Unke and Meuwly(2019)Unke, and Meuwly]{PhysNet}
Unke,~O.~T.; Meuwly,~M. PhysNet: A Neural Network for Predicting Energies, Forces, Dipole Moments, and Partial Charges. \emph{J. Chem. Theory Comput.} \textbf{2019}, \emph{15}, 3678--3693\relax
\mciteBstWouldAddEndPuncttrue
\mciteSetBstMidEndSepPunct{\mcitedefaultmidpunct}
{\mcitedefaultendpunct}{\mcitedefaultseppunct}\relax
\EndOfBibitem
\bibitem[Dral \latin{et~al.}(2017)Dral, Owens, Yurchenko, and Thiel]{KREG}
Dral,~P.~O.; Owens,~A.; Yurchenko,~S.~N.; Thiel,~W. Structure-based sampling and self-correcting machine learning for accurate calculations of potential energy surfaces and vibrational levels. \emph{J. Chem. Phys.} \textbf{2017}, \emph{146}, 244108\relax
\mciteBstWouldAddEndPuncttrue
\mciteSetBstMidEndSepPunct{\mcitedefaultmidpunct}
{\mcitedefaultendpunct}{\mcitedefaultseppunct}\relax
\EndOfBibitem
\bibitem[Dral(2019)]{pKREG}
Dral,~P.~O. MLatom: A program package for quantum chemical research assisted by machine learning. \emph{J. Comput. Chem} \textbf{2019}, \emph{40}, 2339--2347\relax
\mciteBstWouldAddEndPuncttrue
\mciteSetBstMidEndSepPunct{\mcitedefaultmidpunct}
{\mcitedefaultendpunct}{\mcitedefaultseppunct}\relax
\EndOfBibitem
\bibitem[Qu and Bowman(2016)Qu, and Bowman]{Qu2016}
Qu,~C.; Bowman,~J.~M. {An ab initio potential energy surface for the formic acid dimer: Zero-point energy, selected anharmonic fundamental energies, and ground-state tunneling splitting calculated in relaxed 1--4-mode subspaces}. \emph{Phys. Chem. Chem. Phys.} \textbf{2016}, \emph{18}, 24835--24840\relax
\mciteBstWouldAddEndPuncttrue
\mciteSetBstMidEndSepPunct{\mcitedefaultmidpunct}
{\mcitedefaultendpunct}{\mcitedefaultseppunct}\relax
\EndOfBibitem
\bibitem[Rasheeda \latin{et~al.}(2022)Rasheeda, Santa~Dar{\'\i}a, Schr{\"o}der, M{\'a}tyus, and Behler]{fad2022}
Rasheeda,~D.~S.; Santa~Dar{\'\i}a,~A.~M.; Schr{\"o}der,~B.; M{\'a}tyus,~E.; Behler,~J. High-dimensional neural network potentials for accurate vibrational frequencies: the formic acid dimer benchmark. \emph{Phys. Chem. Chem. Phys.} \textbf{2022}, \emph{24}, 29381--29392\relax
\mciteBstWouldAddEndPuncttrue
\mciteSetBstMidEndSepPunct{\mcitedefaultmidpunct}
{\mcitedefaultendpunct}{\mcitedefaultseppunct}\relax
\EndOfBibitem
\bibitem[Fu \latin{et~al.}(2020)Fu, Lu, Han, Fu, Zhang, and Bowman]{Furoam20}
Fu,~Y.-L.; Lu,~X.; Han,~Y.-C.; Fu,~B.; Zhang,~D.~H.; Bowman,~J.~M. Collision-induced and complex-mediated roaming dynamics in the \ce{H} + \ce{C2H4} $\rightarrow$ \ce{H2 + C2H3} reaction. \emph{Chem. Sci.} \textbf{2020}, \emph{11}, 2148--2154\relax
\mciteBstWouldAddEndPuncttrue
\mciteSetBstMidEndSepPunct{\mcitedefaultmidpunct}
{\mcitedefaultendpunct}{\mcitedefaultseppunct}\relax
\EndOfBibitem
\bibitem[Lu \latin{et~al.}(2020)Lu, Behler, and Li]{HCH3OH}
Lu,~D.; Behler,~J.; Li,~J. Accurate Global Potential Energy Surfaces for the \ce{H} + \ce{CH3OH} Reaction by Neural Network Fitting with Permutation Invariance. \emph{J. Phys. Chem. A} \textbf{2020}, \emph{124}, 5737--5745\relax
\mciteBstWouldAddEndPuncttrue
\mciteSetBstMidEndSepPunct{\mcitedefaultmidpunct}
{\mcitedefaultendpunct}{\mcitedefaultseppunct}\relax
\EndOfBibitem
\bibitem[Papp \latin{et~al.}(2020)Papp, Tajti, Győri, and Czakó]{cazko9atom}
Papp,~D.; Tajti,~V.; Győri,~T.; Czakó,~G. Theory Finally Agrees with Experiment for the Dynamics of the \ce{Cl + C2H6} Reaction. \emph{J. Phys. Chem. Lett.} \textbf{2020}, \emph{11}, 4762--4767\relax
\mciteBstWouldAddEndPuncttrue
\mciteSetBstMidEndSepPunct{\mcitedefaultmidpunct}
{\mcitedefaultendpunct}{\mcitedefaultseppunct}\relax
\EndOfBibitem
\bibitem[Czakó \latin{et~al.}(2024)Czakó, Gruber, Papp, Tajti, Tasi, and Yin]{czako2024}
Czakó,~G.; Gruber,~B.; Papp,~D.; Tajti,~V.; Tasi,~D.~A.; Yin,~C. First-principles mode-specific reaction dynamics. \emph{Phys. Chem. Chem. Phys.} \textbf{2024}, \emph{26}, 15818--15830\relax
\mciteBstWouldAddEndPuncttrue
\mciteSetBstMidEndSepPunct{\mcitedefaultmidpunct}
{\mcitedefaultendpunct}{\mcitedefaultseppunct}\relax
\EndOfBibitem
\bibitem[Hamilton \latin{et~al.}(1986)Hamilton, Light, and Whaley]{Light1986}
Hamilton,~I.~P.; Light,~J.~C.; Whaley,~K.~B. Potential energy determination by inverse perturbation analysis with local correction functions. \emph{J. Chem. Phys.} \textbf{1986}, \emph{85}, 5151--5157\relax
\mciteBstWouldAddEndPuncttrue
\mciteSetBstMidEndSepPunct{\mcitedefaultmidpunct}
{\mcitedefaultendpunct}{\mcitedefaultseppunct}\relax
\EndOfBibitem
\bibitem[Wu and Zhang(1996)Wu, and Zhang]{WU1996}
Wu,~Q.; Zhang,~J. Z.~H. Correction of potential energy surface using inverse perturbation and singular value decomposition. \emph{Chem. Phys. Letts} \textbf{1996}, \emph{252}, 195--200\relax
\mciteBstWouldAddEndPuncttrue
\mciteSetBstMidEndSepPunct{\mcitedefaultmidpunct}
{\mcitedefaultendpunct}{\mcitedefaultseppunct}\relax
\EndOfBibitem
\bibitem[Skokov \latin{et~al.}(1999)Skokov, Peterson, and Bowman]{SKOKOV99}
Skokov,~S.; Peterson,~K.~A.; Bowman,~J.~M. Perturbative inversion of the \ce{HOCl} potential energy surface via singular value decomposition. \emph{Chem. Phys. Letts} \textbf{1999}, \emph{312}, 494 -- 502\relax
\mciteBstWouldAddEndPuncttrue
\mciteSetBstMidEndSepPunct{\mcitedefaultmidpunct}
{\mcitedefaultendpunct}{\mcitedefaultseppunct}\relax
\EndOfBibitem
\bibitem[Gazdy and Bowman(1991)Gazdy, and Bowman]{gbhcnmorph}
Gazdy,~B.; Bowman,~J.~M. An adjusted global potential surface for HCN based on rigorous vibrational calculations. \emph{J. Chem. Phys.} \textbf{1991}, \emph{95}, 6309--6316\relax
\mciteBstWouldAddEndPuncttrue
\mciteSetBstMidEndSepPunct{\mcitedefaultmidpunct}
{\mcitedefaultendpunct}{\mcitedefaultseppunct}\relax
\EndOfBibitem
\bibitem[Bowman and Gazdy(1991)Bowman, and Gazdy]{gbscaling}
Bowman,~J.~M.; Gazdy,~B. A simple method to adjust potential energy surfaces: Application to \ce{HCO}. \emph{J. Chem. Phys.} \textbf{1991}, \emph{94}, 816--817\relax
\mciteBstWouldAddEndPuncttrue
\mciteSetBstMidEndSepPunct{\mcitedefaultmidpunct}
{\mcitedefaultendpunct}{\mcitedefaultseppunct}\relax
\EndOfBibitem
\bibitem[Meuwly and Hutson(1999)Meuwly, and Hutson]{meuwlymorph}
Meuwly,~M.; Hutson,~J.~M. Morphing ab initio potentials: A systematic study of \ce{Ne-HF}. \emph{J. Chem. Phys.} \textbf{1999}, \emph{110}, 8338--8347\relax
\mciteBstWouldAddEndPuncttrue
\mciteSetBstMidEndSepPunct{\mcitedefaultmidpunct}
{\mcitedefaultendpunct}{\mcitedefaultseppunct}\relax
\EndOfBibitem
\bibitem[Pan and Yang(2010)Pan, and Yang]{TL_ieee}
Pan,~S.~J.; Yang,~Q. A Survey on Transfer Learning. \emph{IEEE Trans. Knowl. Data Eng.} \textbf{2010}, \emph{22}, 1345--1359\relax
\mciteBstWouldAddEndPuncttrue
\mciteSetBstMidEndSepPunct{\mcitedefaultmidpunct}
{\mcitedefaultendpunct}{\mcitedefaultseppunct}\relax
\EndOfBibitem
\bibitem[Ramakrishnan \latin{et~al.}(2015)Ramakrishnan, Dral, Rupp, and von Lilienfeld]{Lilienfeld15}
Ramakrishnan,~R.; Dral,~P.~O.; Rupp,~M.; von Lilienfeld,~O.~A. Big Data Meets Quantum Chemistry Approximations: The $\Delta$-Machine Learning Approach. \emph{J. Chem. Theory Comput.} \textbf{2015}, \emph{11}, 2087--2096\relax
\mciteBstWouldAddEndPuncttrue
\mciteSetBstMidEndSepPunct{\mcitedefaultmidpunct}
{\mcitedefaultendpunct}{\mcitedefaultseppunct}\relax
\EndOfBibitem
\bibitem[Zaspel \latin{et~al.}(2019)Zaspel, Huang, Harbrecht, and von Lilienfeld]{Lilienfeld19}
Zaspel,~P.; Huang,~B.; Harbrecht,~H.; von Lilienfeld,~O.~A. Boosting Quantum Machine Learning Models with a Multilevel Combination Technique: Pople Diagrams Revisited. \emph{J. Chem. Theory and Comput.} \textbf{2019}, \emph{15}, 1546--1559\relax
\mciteBstWouldAddEndPuncttrue
\mciteSetBstMidEndSepPunct{\mcitedefaultmidpunct}
{\mcitedefaultendpunct}{\mcitedefaultseppunct}\relax
\EndOfBibitem
\bibitem[St{\"o}hr \latin{et~al.}(2020)St{\"o}hr, Medrano~Sandonas, and Tkatchenko]{Stohr2020}
St{\"o}hr,~M.; Medrano~Sandonas,~L.; Tkatchenko,~A. Accurate Many-Body Repulsive Potentials for Density-Functional Tight Binding from Deep Tensor Neural Networks. \emph{J. Phys. Chem. Lett.} \textbf{2020}, \emph{11}, 6835--6843\relax
\mciteBstWouldAddEndPuncttrue
\mciteSetBstMidEndSepPunct{\mcitedefaultmidpunct}
{\mcitedefaultendpunct}{\mcitedefaultseppunct}\relax
\EndOfBibitem
\bibitem[Dral \latin{et~al.}(2020)Dral, Owens, Dral, and Cs\'anyi]{Csanyi_DeltaML}
Dral,~P.~O.; Owens,~A.; Dral,~A.; Cs\'anyi,~G. Hierarchical machine learning of potential energy surfaces. \emph{J. Chem. Phys.} \textbf{2020}, \emph{152}, 204110\relax
\mciteBstWouldAddEndPuncttrue
\mciteSetBstMidEndSepPunct{\mcitedefaultmidpunct}
{\mcitedefaultendpunct}{\mcitedefaultseppunct}\relax
\EndOfBibitem
\bibitem[Käser \latin{et~al.}(2022)Käser, Richardson, and Meuwly]{meuwly2022}
Käser,~S.; Richardson,~J.~O.; Meuwly,~M. Transfer Learning for Affordable and High-Quality Tunneling Splittings from Instanton Calculations. \emph{J. Chem. Theory Comput.} \textbf{2022}, \emph{18}, 6840--6850\relax
\mciteBstWouldAddEndPuncttrue
\mciteSetBstMidEndSepPunct{\mcitedefaultmidpunct}
{\mcitedefaultendpunct}{\mcitedefaultseppunct}\relax
\EndOfBibitem
\bibitem[Nandi \latin{et~al.}(2021)Nandi, Qu, Houston, Conte, and Bowman]{Nandi_Bowman_DeltaML}
Nandi,~A.; Qu,~C.; Houston,~P.~L.; Conte,~R.; Bowman,~J.~M. $\Delta$-machine learning for potential energy surfaces: A PIP approach to bring a DFT-based PES to CCSD(T) level of theory. \emph{J. Chem. Phys.} \textbf{2021}, \emph{154}, 051102\relax
\mciteBstWouldAddEndPuncttrue
\mciteSetBstMidEndSepPunct{\mcitedefaultmidpunct}
{\mcitedefaultendpunct}{\mcitedefaultseppunct}\relax
\EndOfBibitem
\bibitem[Qu \latin{et~al.}(2021)Qu, Houston, Conte, Nandi, and Bowman]{acac2021}
Qu,~C.; Houston,~P.~L.; Conte,~R.; Nandi,~A.; Bowman,~J.~M. Breaking the Coupled Cluster Barrier for Machine-Learned Potentials of Large Molecules: The Case of 15-Atom Acetylacetone. \emph{J. Phys. Chem. Letts.} \textbf{2021}, \emph{12}, 4902--4909\relax
\mciteBstWouldAddEndPuncttrue
\mciteSetBstMidEndSepPunct{\mcitedefaultmidpunct}
{\mcitedefaultendpunct}{\mcitedefaultseppunct}\relax
\EndOfBibitem
\bibitem[Khire \latin{et~al.}(2022)Khire, Gurav, Nandi, and Gadre]{AcAc_MTA}
Khire,~S.~S.; Gurav,~N.~D.; Nandi,~A.; Gadre,~S.~R. Enabling Rapid and Accurate Construction of CCSD(T)-Level Potential Energy Surface of Large Molecules Using Molecular Tailoring Approach. \emph{J. Phys. Chem. A} \textbf{2022}, \emph{126}, 1458--1464\relax
\mciteBstWouldAddEndPuncttrue
\mciteSetBstMidEndSepPunct{\mcitedefaultmidpunct}
{\mcitedefaultendpunct}{\mcitedefaultseppunct}\relax
\EndOfBibitem
\bibitem[Nandi and Nagy(2024)Nandi, and Nagy]{NANDI_AICHEM}
Nandi,~A.; Nagy,~P.~R. Combining state-of-the-art quantum chemistry and machine learning make gold standard potential energy surfaces accessible for medium-sized molecules. \emph{Artif. Intell. Chem.} \textbf{2024}, \emph{2}, 100036\relax
\mciteBstWouldAddEndPuncttrue
\mciteSetBstMidEndSepPunct{\mcitedefaultmidpunct}
{\mcitedefaultendpunct}{\mcitedefaultseppunct}\relax
\EndOfBibitem
\bibitem[Nandi \latin{et~al.}(2022)Nandi, Conte, Qu, Houston, Yu, and Bowman]{nandi2022quantum}
Nandi,~A.; Conte,~R.; Qu,~C.; Houston,~P.~L.; Yu,~Q.; Bowman,~J.~M. Quantum calculations on a new CCSD (T) machine-learned potential energy surface reveal the leaky nature of gas-phase trans and gauche ethanol conformers. \emph{J. Chem. Theory Comput.} \textbf{2022}, \emph{18}, 5527--5538\relax
\mciteBstWouldAddEndPuncttrue
\mciteSetBstMidEndSepPunct{\mcitedefaultmidpunct}
{\mcitedefaultendpunct}{\mcitedefaultseppunct}\relax
\EndOfBibitem
\bibitem[Conte \latin{et~al.}(2022)Conte, Nandi, Qu, Yu, Houston, and Bowman]{ethanolMM}
Conte,~R.; Nandi,~A.; Qu,~C.; Yu,~Q.; Houston,~P.~L.; Bowman,~J.~M. Semiclassical and VSCF/VCI Calculations of the Vibrational Energies of trans- and gauche-Ethanol Using a CCSD(T) Potential Energy Surface. \emph{J. Phys. Chem. A} \textbf{2022}, \emph{126}, 7709--7718\relax
\mciteBstWouldAddEndPuncttrue
\mciteSetBstMidEndSepPunct{\mcitedefaultmidpunct}
{\mcitedefaultendpunct}{\mcitedefaultseppunct}\relax
\EndOfBibitem
\bibitem[Nandi \latin{et~al.}(2023)Nandi, Laude, Khire, Gurav, Qu, Conte, Yu, Li, Houston, Gadre, Richardson, Evangelista, and Bowman]{Nandi2023JACS}
Nandi,~A.; Laude,~G.; Khire,~S.~S.; Gurav,~N.~D.; Qu,~C.; Conte,~R.; Yu,~Q.; Li,~S.; Houston,~P.~L.; Gadre,~S.~R.; Richardson,~J.~O.; Evangelista,~F.~A.; Bowman,~J.~M. Ring-Polymer Instanton Tunneling Splittings of Tropolone and Isotopomers using a $\Delta$-Machine Learned CCSD(T) Potential: Theory and Experiment Shake Hands. \emph{J. Amer. Chem. Soc.} \textbf{2023}, \emph{145}, 9655--9664\relax
\mciteBstWouldAddEndPuncttrue
\mciteSetBstMidEndSepPunct{\mcitedefaultmidpunct}
{\mcitedefaultendpunct}{\mcitedefaultseppunct}\relax
\EndOfBibitem
\bibitem[Houston \latin{et~al.}(2024)Houston, Qu, Yu, Pandey, Conte, Nandi, Bowman, and Kukolich]{fanh32024}
Houston,~P.~L.; Qu,~C.; Yu,~Q.; Pandey,~P.; Conte,~R.; Nandi,~A.; Bowman,~J.~M.; Kukolich,~S.~G. Formic Acid–Ammonia Heterodimer: A New $\Delta$-Machine Learning CCSD(T)-Level Potential Energy Surface Allows Investigation of the Double Proton Transfer. \emph{J. Chem. Theo. Comp.} \textbf{2024}, \emph{20}, 1821--1828\relax
\mciteBstWouldAddEndPuncttrue
\mciteSetBstMidEndSepPunct{\mcitedefaultmidpunct}
{\mcitedefaultendpunct}{\mcitedefaultseppunct}\relax
\EndOfBibitem
\bibitem[Bowman \latin{et~al.}(2023)Bowman, Qu, Conte, Nandi, Houston, and Yu]{JCTCPers2023}
Bowman,~J.~M.; Qu,~C.; Conte,~R.; Nandi,~A.; Houston,~P.~L.; Yu,~Q. $\Delta$-Machine Learned Potential Energy Surfaces and Force Fields. \emph{J. Chem. Theory Comput.} \textbf{2023}, \emph{19}, 1--17\relax
\mciteBstWouldAddEndPuncttrue
\mciteSetBstMidEndSepPunct{\mcitedefaultmidpunct}
{\mcitedefaultendpunct}{\mcitedefaultseppunct}\relax
\EndOfBibitem
\bibitem[Lee \latin{et~al.}(1988)Lee, Yang, and Parr]{LWP}
Lee,~C.; Yang,~W.; Parr,~R.~G. Development of the Colle-Salvetti correlation-energy formula into a functional of the electron density. \emph{Phys. Rev. B} \textbf{1988}, \emph{37}, 785--789\relax
\mciteBstWouldAddEndPuncttrue
\mciteSetBstMidEndSepPunct{\mcitedefaultmidpunct}
{\mcitedefaultendpunct}{\mcitedefaultseppunct}\relax
\EndOfBibitem
\bibitem[Becke(1993)]{becke93}
Becke,~A.~D. {Density‐functional thermochemistry. III. The role of exact exchange}. \emph{J. Chem. Phys.} \textbf{1993}, \emph{98}, 5648--5652\relax
\mciteBstWouldAddEndPuncttrue
\mciteSetBstMidEndSepPunct{\mcitedefaultmidpunct}
{\mcitedefaultendpunct}{\mcitedefaultseppunct}\relax
\EndOfBibitem
\bibitem[Devereux \latin{et~al.}(2024)Devereux, Boittier, and Meuwly]{meuwlyFFDelta}
Devereux,~M.; Boittier,~E.~D.; Meuwly,~M. Systematic improvement of empirical energy functions in the era of machine learning. \emph{J. Comp. Chem.} \textbf{2024}, \emph{45}, 1899--1913\relax
\mciteBstWouldAddEndPuncttrue
\mciteSetBstMidEndSepPunct{\mcitedefaultmidpunct}
{\mcitedefaultendpunct}{\mcitedefaultseppunct}\relax
\EndOfBibitem
\bibitem[Unke \latin{et~al.}(2024)Unke, Stöhr, Ganscha, Unterthiner, Maennel, Kashubin, Ahlin, Gastegger, Sandonas, Berryman, Tkatchenko, and Müller]{sciadv2024}
Unke,~O.~T.; Stöhr,~M.; Ganscha,~S.; Unterthiner,~T.; Maennel,~H.; Kashubin,~S.; Ahlin,~D.; Gastegger,~M.; Sandonas,~L.~M.; Berryman,~J.~T.; Tkatchenko,~A.; Müller,~K.-R. Biomolecular dynamics with machine-learned quantum-mechanical force fields trained on diverse chemical fragments. \emph{Sci. Adv.} \textbf{2024}, \emph{10}, eadn4397\relax
\mciteBstWouldAddEndPuncttrue
\mciteSetBstMidEndSepPunct{\mcitedefaultmidpunct}
{\mcitedefaultendpunct}{\mcitedefaultseppunct}\relax
\EndOfBibitem
\bibitem[Plé \latin{et~al.}(2023)Plé, Lagardère, and Piquemal]{D3SC02581K}
Plé,~T.; Lagardère,~L.; Piquemal,~J.-P. Force-field-enhanced neural network interactions: from local equivariant embedding to atom-in-molecule properties and long-range effects. \emph{Chem. Sci.} \textbf{2023}, \emph{14}, 12554--12569\relax
\mciteBstWouldAddEndPuncttrue
\mciteSetBstMidEndSepPunct{\mcitedefaultmidpunct}
{\mcitedefaultendpunct}{\mcitedefaultseppunct}\relax
\EndOfBibitem
\bibitem[Song \latin{et~al.}(2023)Song, Käser, Töpfer, Vazquez-Salazar, and Meuwly]{Meuwly2023}
Song,~K.; Käser,~S.; Töpfer,~K.; Vazquez-Salazar,~L.~I.; Meuwly,~M. {PhysNet meets CHARMM: A framework for routine machine learning/molecular mechanics simulations}. \emph{J. Chem. Phys.} \textbf{2023}, \emph{159}, 024125\relax
\mciteBstWouldAddEndPuncttrue
\mciteSetBstMidEndSepPunct{\mcitedefaultmidpunct}
{\mcitedefaultendpunct}{\mcitedefaultseppunct}\relax
\EndOfBibitem
\bibitem[Rogers and Wang(2020)Rogers, and Wang]{rogers2020accurate}
Rogers,~T.~R.; Wang,~F. Accurate MP2-based force fields predict hydration free energies for simple alkanes and alcohols in good agreement with experiments. \emph{J. Chem. Phys.} \textbf{2020}, \emph{153}, 244505\relax
\mciteBstWouldAddEndPuncttrue
\mciteSetBstMidEndSepPunct{\mcitedefaultmidpunct}
{\mcitedefaultendpunct}{\mcitedefaultseppunct}\relax
\EndOfBibitem
\bibitem[Zhao and Truhlar(2008)Zhao, and Truhlar]{MO6}
Zhao,~Y.; Truhlar,~D.~G. The M06 suite of density functionals for main group thermochemistry, thermochemical kinetics, noncovalent interactions, excited states, and transition elements: two new functionals and systematic testing of four M06-class functionals and 12 other functionals. \emph{Theor. Chem. Acc.} \textbf{2008}, \emph{120}, 215--241\relax
\mciteBstWouldAddEndPuncttrue
\mciteSetBstMidEndSepPunct{\mcitedefaultmidpunct}
{\mcitedefaultendpunct}{\mcitedefaultseppunct}\relax
\EndOfBibitem
\bibitem[Walker \latin{et~al.}(2013)Walker, Harvey, Sen, and Dessent]{M06_2X}
Walker,~M.; Harvey,~A. J.~A.; Sen,~A.; Dessent,~C. E.~H. Performance of M06, M06-2X, and M06-HF Density Functionals for Conformationally Flexible Anionic Clusters: M06 Functionals Perform Better than B3LYP for a Model System with Dispersion and Ionic Hydrogen-Bonding Interactions. \emph{J. Phys. Chem. A} \textbf{2013}, \emph{117}, 12590--12600\relax
\mciteBstWouldAddEndPuncttrue
\mciteSetBstMidEndSepPunct{\mcitedefaultmidpunct}
{\mcitedefaultendpunct}{\mcitedefaultseppunct}\relax
\EndOfBibitem
\bibitem[Perdew \latin{et~al.}(1996)Perdew, Burke, and Ernzerhof]{PBE96}
Perdew,~J.~P.; Burke,~K.; Ernzerhof,~M. {Generalized Gradient Approximation Made Simple}. \emph{Phys. Rev. Lett.} \textbf{1996}, \emph{77}, 3865--3868, Erratum \textit{Phys. Rev. Lett.} \textbf{1997}, \textit{78}, 1396.\relax
\mciteBstWouldAddEndPunctfalse
\mciteSetBstMidEndSepPunct{\mcitedefaultmidpunct}
{}{\mcitedefaultseppunct}\relax
\EndOfBibitem
\bibitem[Perdew \latin{et~al.}(1996)Perdew, Ernzerhof, and Burke]{PBE0_1}
Perdew,~J.~P.; Ernzerhof,~M.; Burke,~K. {Rationale for mixing exact exchange with density functional approximations}. \emph{J. Chem. Phys.} \textbf{1996}, \emph{105}, 9982--9985\relax
\mciteBstWouldAddEndPuncttrue
\mciteSetBstMidEndSepPunct{\mcitedefaultmidpunct}
{\mcitedefaultendpunct}{\mcitedefaultseppunct}\relax
\EndOfBibitem
\bibitem[Adamo and Barone(1999)Adamo, and Barone]{PBE0_2}
Adamo,~C.; Barone,~V. {Toward reliable density functional methods without adjustable parameters: The PBE0 model}. \emph{J. Chem. Phys.} \textbf{1999}, \emph{110}, 6158--6170\relax
\mciteBstWouldAddEndPuncttrue
\mciteSetBstMidEndSepPunct{\mcitedefaultmidpunct}
{\mcitedefaultendpunct}{\mcitedefaultseppunct}\relax
\EndOfBibitem
\bibitem[Tkatchenko \latin{et~al.}(2012)Tkatchenko, DiStasio, Car, and Scheffler]{mbd2012}
Tkatchenko,~A.; DiStasio,~R.~A.; Car,~R.; Scheffler,~M. Accurate and Efficient Method for Many-Body van der Waals Interactions. \emph{Phys. Rev. Lett.} \textbf{2012}, \emph{108}, 236402\relax
\mciteBstWouldAddEndPuncttrue
\mciteSetBstMidEndSepPunct{\mcitedefaultmidpunct}
{\mcitedefaultendpunct}{\mcitedefaultseppunct}\relax
\EndOfBibitem
\bibitem[Blum \latin{et~al.}(2009)Blum, Gehrke, Hanke, Havu, Havu, Ren, Reuter, and Scheffler]{FHI_aims}
Blum,~V.; Gehrke,~R.; Hanke,~F.; Havu,~P.; Havu,~V.; Ren,~X.; Reuter,~K.; Scheffler,~M. Ab initio molecular simulations with numeric atom-centered orbitals. \emph{Comput. Phys. Commun.} \textbf{2009}, \emph{180}, 2175--2196\relax
\mciteBstWouldAddEndPuncttrue
\mciteSetBstMidEndSepPunct{\mcitedefaultmidpunct}
{\mcitedefaultendpunct}{\mcitedefaultseppunct}\relax
\EndOfBibitem
\bibitem[Braams and Bowman(2009)Braams, and Bowman]{Braams2009}
Braams,~B.~J.; Bowman,~J.~M. Permutationally invariant potential energy surfaces in high dimensionality. \emph{Int. Rev. Phys. Chem.} \textbf{2009}, \emph{28}, 577--606\relax
\mciteBstWouldAddEndPuncttrue
\mciteSetBstMidEndSepPunct{\mcitedefaultmidpunct}
{\mcitedefaultendpunct}{\mcitedefaultseppunct}\relax
\EndOfBibitem
\bibitem[Bowman \latin{et~al.}(2010)Bowman, Braams, Carter, Chen, Cza{\'k}o, Fu, Huang, Kamarchik, Sharma, Shepler, Wang, and Xie]{Bowman2010}
Bowman,~J.~M.; Braams,~B.~J.; Carter,~S.; Chen,~C.; Cza{\'k}o,~G.; Fu,~B.; Huang,~X.; Kamarchik,~E.; Sharma,~A.~R.; Shepler,~B.~C.; Wang,~Y.; Xie,~Z. Ab-initio-based potential energy surfaces for complex molecules and molecular complexes. \emph{J. Phys. Chem. Lett.} \textbf{2010}, \emph{1}, 1866--1874\relax
\mciteBstWouldAddEndPuncttrue
\mciteSetBstMidEndSepPunct{\mcitedefaultmidpunct}
{\mcitedefaultendpunct}{\mcitedefaultseppunct}\relax
\EndOfBibitem
\bibitem[Xie and Bowman(2010)Xie, and Bowman]{Xie10}
Xie,~Z.; Bowman,~J.~M. Permutationally Invariant Polynomial Basis for Molecular Energy Surface Fitting via Monomial Symmetrization. \emph{J. Chem. Theory Comput.} \textbf{2010}, \emph{6}, 26--34\relax
\mciteBstWouldAddEndPuncttrue
\mciteSetBstMidEndSepPunct{\mcitedefaultmidpunct}
{\mcitedefaultendpunct}{\mcitedefaultseppunct}\relax
\EndOfBibitem
\bibitem[Werner \latin{et~al.}(2015)Werner, Knowles, Knizia, Manby, and {Sch\"{u}tz}]{MOLPRO_brief}
Werner,~H.-J.; Knowles,~P.~J.; Knizia,~G.; Manby,~F.~R.; {Sch\"{u}tz},~M. MOLPRO, version 2015.1, a package of ab initio programs. 2015; see http://www.molpro.net\relax
\mciteBstWouldAddEndPuncttrue
\mciteSetBstMidEndSepPunct{\mcitedefaultmidpunct}
{\mcitedefaultendpunct}{\mcitedefaultseppunct}\relax
\EndOfBibitem
\bibitem[Ren \latin{et~al.}(2012)Ren, Rinke, Blum, Wieferink, Tkatchenko, Sanfilippo, Reuter, and Scheffler]{FHI_aims_2}
Ren,~X.; Rinke,~P.; Blum,~V.; Wieferink,~J.; Tkatchenko,~A.; Sanfilippo,~A.; Reuter,~K.; Scheffler,~M. Resolution-of-identity approach to Hartree–Fock, hybrid density functionals, RPA, MP2 and GW with numeric atom-centered orbital basis functions. \emph{New J. Phys.} \textbf{2012}, \emph{14}, 053020\relax
\mciteBstWouldAddEndPuncttrue
\mciteSetBstMidEndSepPunct{\mcitedefaultmidpunct}
{\mcitedefaultendpunct}{\mcitedefaultseppunct}\relax
\EndOfBibitem
\bibitem[msa(2019)]{msachen}
MSA 2.0 Software with Gradients. \url{https://github.com/szquchen/MSA-2.0}, 2019; Accessed: 2019-01-20\relax
\mciteBstWouldAddEndPuncttrue
\mciteSetBstMidEndSepPunct{\mcitedefaultmidpunct}
{\mcitedefaultendpunct}{\mcitedefaultseppunct}\relax
\EndOfBibitem
\bibitem[msa(2024)]{msavideo}
Video MSA 2.0 Software with Gradients. https://scholarblogs.emory.edu/bowman/msa/, 2024\relax
\mciteBstWouldAddEndPuncttrue
\mciteSetBstMidEndSepPunct{\mcitedefaultmidpunct}
{\mcitedefaultendpunct}{\mcitedefaultseppunct}\relax
\EndOfBibitem
\bibitem[Quade(2000)]{Quade2000}
Quade,~C. A Note on Internal Rotation–Rotation Interactions in Ethyl Alcohol. \emph{J. Mol. Spectrosc.} \textbf{2000}, \emph{203}, 200--202\relax
\mciteBstWouldAddEndPuncttrue
\mciteSetBstMidEndSepPunct{\mcitedefaultmidpunct}
{\mcitedefaultendpunct}{\mcitedefaultseppunct}\relax
\EndOfBibitem
\bibitem[Pearson \latin{et~al.}(1995)Pearson, Sastry, Winnewisser, Herbst, and De~Lucia]{Pearson1995}
Pearson,~J.~C.; Sastry,~K. V. L.~N.; Winnewisser,~M.; Herbst,~E.; De~Lucia,~F.~C. The Millimeter‐ and Submillimeter‐Wave Spectrum of trans‐Ethyl Alcohol. \emph{J. Phys. Chem. Ref. Data} \textbf{1995}, \emph{24}, 1--32\relax
\mciteBstWouldAddEndPuncttrue
\mciteSetBstMidEndSepPunct{\mcitedefaultmidpunct}
{\mcitedefaultendpunct}{\mcitedefaultseppunct}\relax
\EndOfBibitem
\bibitem[Durig and Larsen(1990)Durig, and Larsen]{Durig1990}
Durig,~J.; Larsen,~R. Torsional vibrations and barriers to internal rotation for ethanol and 2,2,2-trifluoroethanol. \emph{J. Mol. Struct.} \textbf{1990}, \emph{238}, 195--222\relax
\mciteBstWouldAddEndPuncttrue
\mciteSetBstMidEndSepPunct{\mcitedefaultmidpunct}
{\mcitedefaultendpunct}{\mcitedefaultseppunct}\relax
\EndOfBibitem
\bibitem[Slootman \latin{et~al.}(2024)Slootman, Poltavsky, Shinde, Cocomello, Moroni, Tkatchenko, and Filippi]{slootman2024accurate}
Slootman,~E.; Poltavsky,~I.; Shinde,~R.; Cocomello,~J.; Moroni,~S.; Tkatchenko,~A.; Filippi,~C. Accurate quantum Monte Carlo forces for machine-learned force fields: Ethanol as a benchmark. \emph{J. Chem. Theory Comput.} \textbf{2024}, \relax
\mciteBstWouldAddEndPunctfalse
\mciteSetBstMidEndSepPunct{\mcitedefaultmidpunct}
{}{\mcitedefaultseppunct}\relax
\EndOfBibitem
\end{mcitethebibliography}

\newpage
\section{Supporting Information}

\begin{table}[htbp!]
\centering
\begin{tabular*}{1.0\columnwidth}{@{\extracolsep{\fill}}rrrrrrrrrrrr}
\hline
\hline\noalign{\smallskip}
& CCSD(T)& \multicolumn{2}{c}{PBE} &  \multicolumn{2}{c}{M06} & \multicolumn{2}{c}{M06-2X}& \multicolumn{2}{c}{B3LYP} & \multicolumn{2}{c}{PBE0+MBD} \\
	\noalign{\smallskip} \cline{2-2} \cline{3-4} \cline{5-6} \cline{7-8} \cline{9-10} \cline{11-12} \noalign{\smallskip}
Mode & Direct & $V_{LL}$ & $\Delta$ML & $V_{LL}$ & $\Delta$ML & $V_{LL}$ & $\Delta$ML & $V_{LL}$ & $\Delta$ML & $V_{LL}$ & $\Delta$ML \\
\hline 

1 & 258 & 275 & 273 & 255 & 250 & 271 & 266 & 267& 268 & 261& 257\\
2 & 271 & 305 & 294 & 272 & 269 & 290 & 279 & 279& 278 & 273& 270\\
3 & 420 & 416 & 427 & 422 & 419 & 428 & 423 & 422& 425 & 425& 422\\
4 & 803 & 777 & 805 & 767 & 782 & 803 & 799 & 804& 804 & 800& 805\\
5 & 895 & 875 & 897 & 890 & 888 & 907 & 894 & 882& 895 & 902&894\\
6 & 1069 & 1036 & 1078 & 1054 & 1070 & 1072 &1077 & 1057& 1075 & 1072& 1076\\
7 & 1096 & 1071 & 1100 & 1098 & 1093 & 1112 &1095 & 1075& 1094 & 1106&1093\\
8 & 1141 & 1113 & 1150 & 1137 & 1137 & 1155 &1144 & 1133& 1144 & 1149&1145\\
9 & 1284 & 1237 & 1288 & 1261 & 1283 & 1284 &1288 & 1280& 1290 & 1285&1292\\
10 & 1374 & 1326 & 1374 & 1351 & 1364 &1376 &1374 & 1368& 1375 & 1373&1377\\
11 & 1402 & 1341 & 1406 & 1375 & 1401 &1404 &1405 & 1403& 1406 & 1396&1409\\
12 & 1426 & 1367 & 1418 & 1397 & 1416 &1429 &1425 & 1416& 1424 & 1419&1426\\
13 & 1491 & 1411 & 1487 & 1453 & 1481 &1492 &1489 & 1487& 1490 & 1479&1491\\
14 & 1497 & 1420 & 1493 & 1459 & 1487 &1503 &1499 & 1494& 1496 & 1488&1499\\
15 & 1522 & 1446 & 1516 & 1481 & 1507 &1527 &1521 & 1515& 1519 & 1511&1522\\
16 & 3014 & 2893 & 3013 & 2977 & 2997 &3043 &3006 & 2989& 3007 & 3012&3004\\
17 & 3028 & 2961 & 3034 & 3009 & 3012 &3056 &3016 & 3015& 3020 & 3043&3018\\
18 & 3089 & 3007 & 3109 & 3067 & 3079 &3119 &3081 & 3068& 3089 & 3098& 3085\\
19 & 3108 & 3050 & 3128 & 3095 & 3096 &3137 &3099 & 3087& 3108 & 3124& 3104\\
20 & 3123 & 3069 & 3139 & 3111 & 3108 &3150 &3112 & 3100& 3121 & 3139& 3116\\
21 & 3837 & 3694 & 3850 & 3902 & 3837 &3896 &3831 & 3826& 3845 & 3893& 3841\\
\hline
MAE &   &   55 &    8   &   22  &   8   &   15  &   5   &   11  &   4   &   10  &   4   \\
\hline 
\end{tabular*}
\caption{Comparison of Harmonic Frequencies (in cm$^{-1}$) of Direct DFT and $\Delta$-corrected PES computed using indicated DFT functionals and corresponding Ab Initio (CCSD(T)-F12a/aug-cc-pVDZ) for \textit{gauche}-ethanol}.
\label{tab:hf_pes_gauche}
\end{table}

\begin{table}[htbp!]
\centering
\begin{tabular*}{1.0\columnwidth}{@{\extracolsep{\fill}}rrrrrrrrrrrr}
\hline
\hline\noalign{\smallskip}
& CCSD(T)& \multicolumn{2}{c}{PBE} &  \multicolumn{2}{c}{M06} & \multicolumn{2}{c}{M06-2X}& \multicolumn{2}{c}{B3LYP} & \multicolumn{2}{c}{PBE0+MBD} \\
	\noalign{\smallskip} \cline{2-2} \cline{3-4} \cline{5-6} \cline{7-8} \cline{9-10} \cline{11-12} \noalign{\smallskip}
Mode & Direct & $V_{LL}$ & $\Delta$ML & $V_{LL}$ & $\Delta$ML & $V_{LL}$ & $\Delta$ML & $V_{LL}$ & $\Delta$ML & $V_{LL}$ & $\Delta$ML \\
\hline 

1 & 287i & 289i &  292i & 271i & 266i & 274i & 264i & 261i & 267i & 268i& 263i\\
2 & 256  & 267 & 265 & 258 & 255 & 265 & 261 & 259 & 261 & 260& 259\\
3 & 416 &  409 & 418 & 421 & 418 & 426 & 420 & 420 & 420 & 423&420\\
4 & 797 &  773 & 797 & 766 & 780 & 799 & 796 & 801 & 800 & 798&803\\
5 & 899 &  880 & 901 & 896 & 894 & 913 & 900 & 887 & 899 & 907&900\\
6 & 1064 & 1015 & 1063 & 1035 & 1046 & 1062 & 1058 & 1036 & 1058 & 1060&1059\\
7 & 1106 & 1081 & 1109 & 1101 & 1097 & 1118 & 1103 & 1087 & 1106 & 1117 &1107\\
8 & 1132 & 1097 & 1134 & 1125 & 1128 & 1143 & 1133 & 1126 & 1133 & 1136 &1135\\
9 & 1285 & 1226 & 1281 & 1251 & 1273 & 1281 & 1283 & 1275 & 1285 & 1282 &1288\\
10 & 1358 & 1320 & 1371 & 1342 &1355 & 1370 & 1370 & 1360 & 1370 & 1369 &1372\\
11 & 1397 & 1328 & 1396 & 1373 &1395 & 1398 & 1400 & 1397 & 1399 & 1390 & 1402\\
12 & 1427 & 1382 & 1424 & 1407 &1417 & 1435 & 1425 & 1423 & 1427 & 1426 & 1431\\
13 & 1486 & 1405 & 1484 & 1450 &1479 & 1486 & 1485 & 1481 & 1485 & 1473 & 1486\\
14 & 1498 & 1426 & 1492 & 1462 &1488 & 1507 & 1503 & 1498 & 1500 & 1491 &1502\\
15 & 1520 & 1448 & 1520 & 1486 &1508 & 1531 & 1522 & 1517 & 1522 & 1513 &1523\\
16 & 3028 & 2909 & 3026 & 2988 &3009 & 3052 & 3016 & 3003 & 3020 & 3023 &3016\\
17 & 3034 & 2952 & 3042 & 3017 &3020 & 3064 & 3023 & 3023 & 3028 & 3051 &3025\\
18 & 3069 & 2971 & 3075 & 3030 &3051 & 3088 & 3052 & 3035 & 3059 & 3061 &3056\\
19 & 3123 & 3067 & 3130 & 3106 &3101 & 3144 & 3105 & 3091 & 3112 & 3135 &3108\\
20 & 3124 & 3070 & 3141 & 3111 &3109 & 3150 & 3114 & 3102 & 3122 & 3141 &3117\\
21 & 3890 & 3757 & 3902 & 3957 &3890 & 3947 & 3883 & 3875 & 3896 & 3945 &3889\\
\hline
MAE &   &   55  &   5   &   23  &   11  &   14  &   7   &   13  &   5   &   11  &   7 \\
\hline 
\end{tabular*}
\caption{Comparison of Harmonic Frequencies (in cm$^{-1}$) of Direct DFT and $\Delta$-corrected PES computed using indicated DFT functionals and corresponding Ab Initio (CCSD(T)-F12a/aug-cc-pVDZ) for \textit{eclipsed}-ethanol}.
\label{tab:hf_pes_anti}
\end{table}

\begin{table}[htbp!]
\centering
\begin{tabular*}{1.0\columnwidth}{@{\extracolsep{\fill}}rrrrrrrrrrrr}
\hline
\hline\noalign{\smallskip}
& CCSD(T)& \multicolumn{2}{c}{PBE} &  \multicolumn{2}{c}{M06} & \multicolumn{2}{c}{M06-2X}& \multicolumn{2}{c}{B3LYP} & \multicolumn{2}{c}{PBE0+MBD} \\
	\noalign{\smallskip} \cline{2-2} \cline{3-4} \cline{5-6} \cline{7-8} \cline{9-10} \cline{11-12} \noalign{\smallskip}
Mode & Direct & $V_{LL}$ & $\Delta$ML & $V_{LL}$ & $\Delta$ML & $V_{LL}$ & $\Delta$ML & $V_{LL}$ & $\Delta$ML & $V_{LL}$ & $\Delta$ML \\
\hline 

1 & 300i & 360i & 345i & 310i & 300i & 336i & 329i & 336i & 332i & 305i & 304i\\
2 & 271 &  281 & 274 & 274 & 271 & 279 & 271 & 270 & 270 & 275 &271\\
3 & 414 &  392 & 409 & 414 & 414 & 417 & 411 & 408 & 411 & 412 &412\\
4 & 807 &  784 & 817 & 784 & 797 & 813 & 810 & 812 & 812 & 808 &813\\
5 & 892 &  875 & 896 & 891 & 888 & 904 & 890 & 878 & 892 & 901 &892\\
6 & 1061 & 1017 & 1054 & 1047 & 1059 & 1060 & 1063 & 1044& 1057 & 1055 &1060\\
7 & 1109 & 1089 & 1112 & 1119 & 1105 & 1130 & 1106 & 1079& 1104 & 1121 &1104\\
8 & 1187 & 1147 & 1196 & 1170 & 1183 & 1189 & 1187 & 1182& 1186 & 1186 &1190\\
9 & 1298 & 1260 & 1301 & 1288 & 1298 & 1304 & 1307 & 1292& 1307 & 1307 &1311\\
10 & 1306 & 1268 & 1310 & 1289 & 1306 &1316 & 1308 & 1306& 1308 & 1310 &1311\\
11 & 1402 & 1333 & 1412 & 1379 & 1404 &1404 & 1404 & 1405& 1406 & 1396 &1409\\
12 & 1440 & 1394 & 1439 & 1423 & 1445 &1451 & 1449 & 1440& 1446 & 1445 &1450\\
13 & 1493 & 1412 & 1496 & 1460 & 1489 &1496 & 1492 & 1491& 1493 & 1483 &1495\\
14 & 1502 & 1432 & 1500 & 1469 & 1497 &1510 & 1508 & 1502& 1507 & 1495 &1507\\
15 & 1539 & 1472 & 1535 & 1500 & 1526 &1544 & 1538 & 1531& 1534 & 1531 &1541 \\
16 & 3027 & 2915 & 3028 & 2993 & 3004 &3054 & 3012 & 3010& 3015 & 3029 &3012\\
17 & 3030 & 2942 & 3037 & 3006 & 3019 &3061 & 3024 & 3013& 3027 & 3040 &3024\\
18 & 3061 & 2958 & 3072 & 3026 & 3046 &3090 & 3050 & 3034& 3054 & 3059 &3053\\
19 & 3106 & 3046 & 3117 & 3091 & 3089 &3133 & 3097 & 3082& 3103 & 3121 &3098\\
20 & 3113 & 3061 & 3127 & 3102 & 3095 &3142 & 3102 & 3088& 3109 & 3131 &3106\\
21 & 3865 & 3728 & 3881 & 3922 & 3862 &3921 & 3858 & 3853& 3872 & 3918 &3868\\
\hline
MAE &   &   57  &   8  &   28  &    7   &   16  &   6   &   12  &   6   &   9   &   5\\ 
\hline 
\end{tabular*}
\caption{Comparison of Harmonic Frequencies (in cm$^{-1}$) of Direct DFT and $\Delta$-corrected PES computed using indicated DFT functionals and corresponding Ab Initio (CCSD(T)-F12a/aug-cc-pVDZ) for \textit{syn}-ethanol}.
\label{tab:hf_pes_syn}
\end{table}

\begin{figure}[htbp!]
    \centering
    \includegraphics[width=1.0\linewidth]{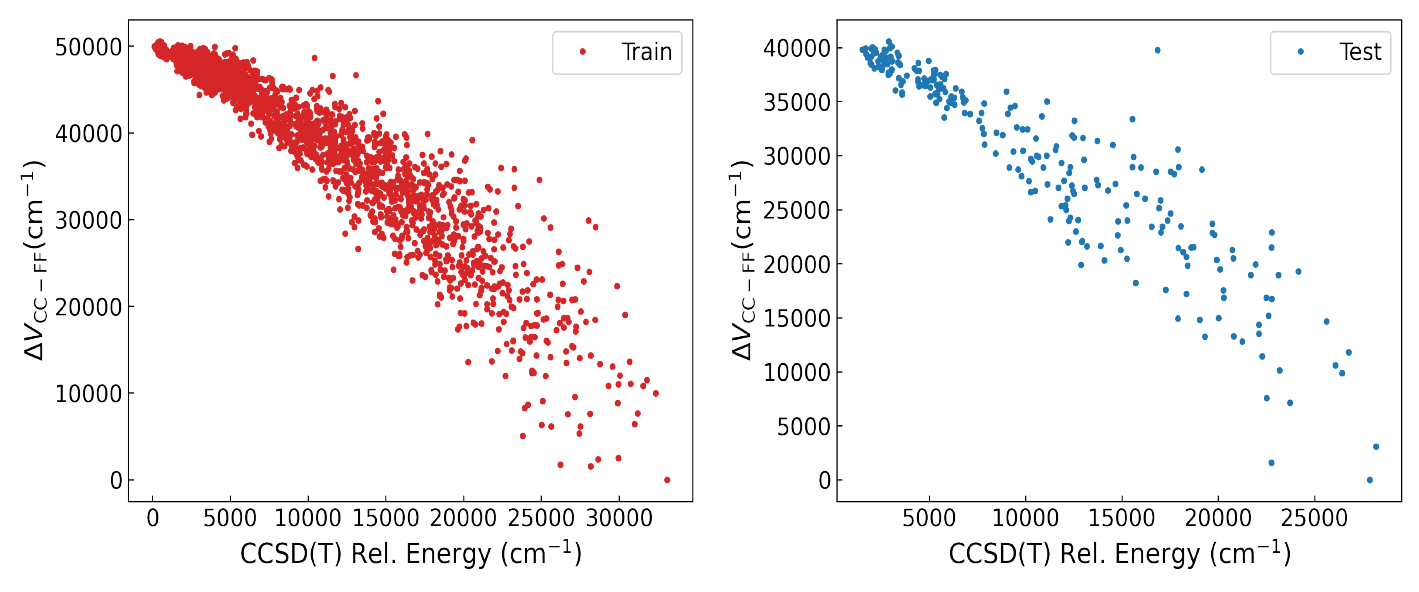}
    \caption{Plot of $\Delta V_{CC-FF}$ vs CCSD(T) energy relative to the \ce{CH3CH2OH} minimum value for the indicated data sets calculated using the harmonic approximation for the MP2 corrected force field.}
    \label{fig:deltaE_ho}
\end{figure}

\begin{figure}[htbp!]
    \centering
    \includegraphics[width=1.0\linewidth]{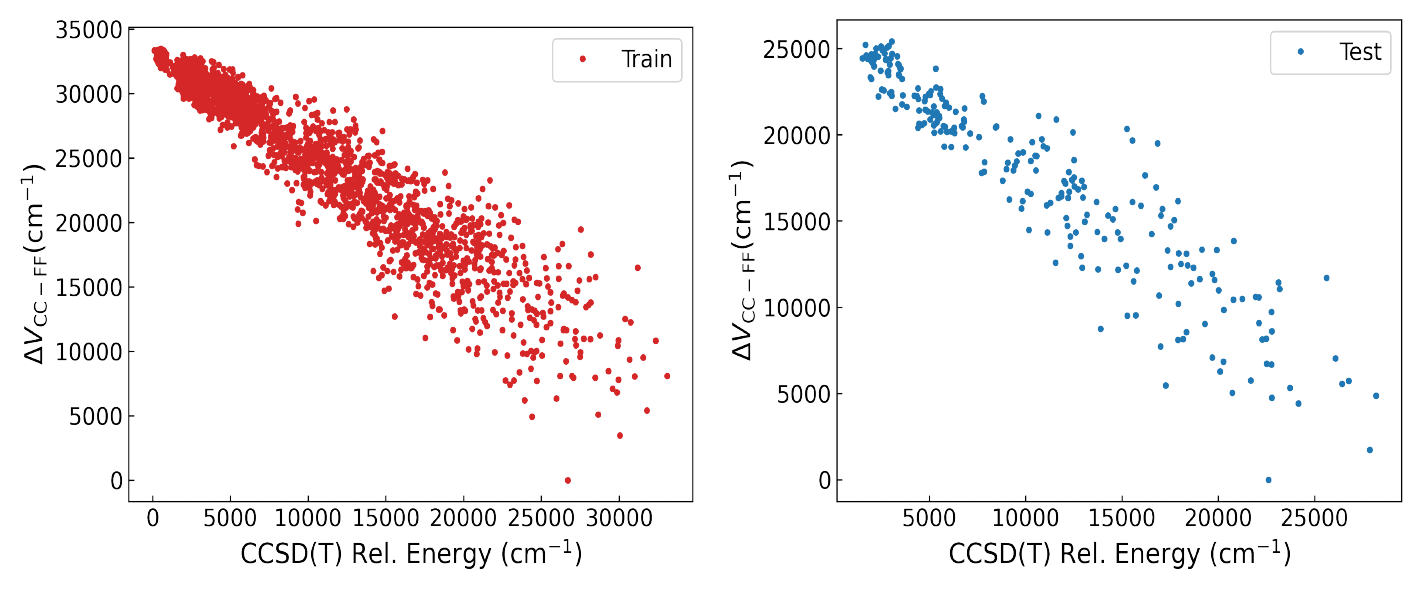}
    \caption{Plot of $\Delta V_{CC-FF}$ vs CCSD(T) energy relative to the \ce{CH3CH2OH} minimum value for the indicated data sets calculated using the Morse potential for the MP2 corrected force field.}
    \label{fig:deltaE_ho}
\end{figure}

\begin{table}[htbp!]
\centering
\begin{tabular*}{1.0\columnwidth}{@{\extracolsep{\fill}}rrrrrrr}
\hline
\hline\noalign{\smallskip}
& CCSD(T)& Force Field & {Harmonic FF} &  \multicolumn{3}{c}{Morse FF}  \\
	\noalign{\smallskip} \cline{2-2}  \cline{3-3}\cline{4-4} \cline{5-7}  \noalign{\smallskip}
Mode & Direct & Harmonic & $\Delta$ML$^a$  & $\Delta$ML$^a$ & $\Delta$ML$^b$ & $\Delta$ML$^c$ \\
\hline 
1 & 258 & 227 & 257 & 309& 222 & 210 \\
2 & 271 & 268 & 302 & 353& 262 & 266 \\
3 & 420 & 606 & 442 & 568& 371 & 426 \\
4 & 803 & 1122& 655 & 700& 657 & 803 \\
5 & 895 & 1186& 779 & 814& 888 & 959 \\
6 & 1069 &1280& 854 & 956& 1028 & 1069 \\
7 & 1096 &1430& 1095& 1043& 1131 & 1152 \\
8 & 1141 &1458& 1212& 1059& 1166 & 1154 \\
9 & 1284 &1724& 1315& 1155& 1172 & 1237 \\
10& 1374 &1821& 1363& 1185& 1270 & 1255 \\
11& 1402 &1961& 1460& 1283& 1328 & 1447 \\
12& 1426 &1993& 1502& 1390& 1394 & 1490 \\
13& 1491 &2019& 1529& 1415& 1411 & 1548 \\
14& 1497 &2044& 1619& 1445& 1455 & 1583 \\
15& 1522 &2146& 1785& 1521& 1773 & 1827\\
16& 3014 &4247& 2005& 3362& 3106 & 3003\\
17& 3028 &4300& 2274& 3448& 3283 & 3138\\
18& 3089 &4399& 2365& 3464& 3345 & 3225\\
19& 3108 &4409& 2501& 3499& 3368 & 3259\\
20& 3123 &4410& 2740& 3513& 3426 & 3353\\
21& 3837 &5129& 3552& 4099& 4137 & 3866\\
\hline 
MAE &     &624 & 236 & 167 & 119 & 75 \\
\hline
\end{tabular*}
\\
         $^a$ Fit using full data points up to 35000 cm$^{-1}$  \\
         $^b$ Fit using data points up to 10000 cm$^{-1}$\\
         $^c$ Fit using data points up to 5000 cm$^{-1}$\\
\caption{Comparison of Harmonic Frequencies (in cm$^{-1}$) between V$_{LL \rightarrow CC}$ PES computed using the force field and corresponding Ab Initio (CCSD(T)-F12a/aug-cc-pVDZ) for $gauche$-ethanol.}
\label{tab:ff_freq_gauche}
\end{table}

\begin{table}[htbp!]
\centering
\begin{tabular*}{1.0\columnwidth}{@{\extracolsep{\fill}}rrrrrrr}
\hline
\hline\noalign{\smallskip}
& CCSD(T)& Force Field & {Harmonic FF} &  \multicolumn{3}{c}{Morse FF}  \\
	\noalign{\smallskip} \cline{2-2}  \cline{3-3}\cline{4-4} \cline{5-7}  \noalign{\smallskip}
Mode & Direct & Harmonic & $\Delta$ML$^a$  & $\Delta$ML$^a$ & $\Delta$ML$^b$ & $\Delta$ML$^c$ \\
\hline 
1 & 287i & 261i & 246i & 363 & 269i & 229i\\
2 & 256 & 233 & 278 & 374 & 243 & 245 \\
3 & 416 & 612 & 422 & 450 & 369 & 414\\
4 & 797 & 1118 & 636 & 734 & 707 & 773 \\
5 & 899 & 1169 & 808 & 854 & 902 & 961\\
6 & 1064 & 1296 & 853 & 973 & 1040 & 1055\\
7 & 1106 & 1418 & 1115 & 1053 & 1116 & 1131\\
8 & 1132 & 1451 & 1226 & 1075 & 1162 & 1171\\
9 & 1285 & 1745 & 1324 & 1149 & 1220 & 1269\\
10& 1358 & 1822 & 1411 & 1192 & 1298 & 1286\\
11& 1397 & 1959 & 1482 & 1264 & 1349 & 1430\\
12& 1427 & 1994 & 1548 & 1325 & 1382 & 1472\\
13& 1486 & 2019 & 1564 & 1390 & 1420 & 1549\\
14& 1498 & 2046 & 1599 & 1476 & 1449 & 1582\\
15& 1520 & 2138 & 1793 & 1480 & 1616 & 1708\\
16& 3028 & 4247 & 1969 & 3356 & 3046 & 3110\\
17& 3034 & 4300 & 2234 & 3443 & 3248 & 3146\\
18& 3069 & 4399 & 2298 & 3471 & 3299 & 3171\\
19& 3123 & 4409 & 2452 & 3480 & 3317 & 3270\\
20& 3124 & 4410 & 2701 & 3503 & 3355 & 3324\\
21& 3890 & 5129 & 3535 & 4101 & 4150 & 3917\\
\hline
MAE &     &623 & 260 & 185 & 86 & 67 \\
\hline
\end{tabular*}
\\
         $^a$ Fit using full data points up to 35000 cm$^{-1}$  \\
         $^b$ Fit using data points up to 10000 cm$^{-1}$\\
         $^c$ Fit using data points up to 5000 cm$^{-1}$\\
\caption{Comparison of Harmonic Frequencies (in cm$^{-1}$) between V$_{LL \rightarrow CC}$ PES computed using the force field and corresponding ab initio (CCSD(T)-F12a/aug-cc-pVDZ) for $eclipsed$-ethanol.}
\label{tab:ff_freq_anti}
\end{table}

\begin{table}[htbp!]
\centering
\begin{tabular*}{1.0\columnwidth}{@{\extracolsep{\fill}}rrrrrrr}
\hline
\hline\noalign{\smallskip}
& CCSD(T)& Force Field & {Harmonic FF} &  \multicolumn{3}{c}{Morse FF}  \\
	\noalign{\smallskip} \cline{2-2}  \cline{3-3}\cline{4-4} \cline{5-7}  \noalign{\smallskip}
Mode & Direct & Harmonic & $\Delta$ML$^a$  & $\Delta$ML$^a$ & $\Delta$ML$^b$ & $\Delta$ML$^c$ \\
\hline 
1 & 300i & 268i & 301i & 274 & 161i & 163i\\
2 & 271 & 232 & 305 & 325 & 252 & 279\\
3 & 414 & 589 & 452 & 528 & 317 & 426\\
4 & 807 & 1130 & 711 & 684 & 646 & 849\\
5 & 892 & 1189 & 735 & 818 & 892 & 971\\
6 & 1061 & 1268 & 816 & 938 & 931 & 994\\
7 & 1109 & 1441 & 1092 & 1042 & 1116 & 1151\\
8 & 1187 & 1457 & 1244 & 1051 & 1121 & 1176\\
9 & 1298 & 1708 & 1290 & 1165 & 1165 & 1213\\
10& 1306 & 1816 & 1343 & 1178 & 1261 & 1252\\
11& 1402 & 1963 & 1417 & 1289 & 1350 & 1523\\
12& 1440 & 1993 & 1481 & 1389 & 1407 & 1536\\
13& 1493 & 2019 & 1483 & 1420 & 1419 & 1635\\
14& 1502 & 2045 & 1637 & 1434 & 1540 & 1658\\
15& 1539 & 2159 & 1772 & 1565 & 1794 & 1887\\
16& 3027 & 4247 & 2060 & 3368 & 3123 & 2931\\
17& 3030 & 4300 & 2302 & 3453 & 3241 & 3149\\
18& 3061 & 4398 & 2416 & 3456 & 3345 & 3260\\
19& 3106 & 4409 & 2544 & 3508 & 3366 & 3272\\
20& 3113 & 4410 & 2770 & 3518 & 3386 & 3365\\
21& 3865 & 5128 & 3560 & 4110 & 4116 & 3896\\
\hline 
MAE &     &623 & 223 & 194 & 125 & 108 \\
\hline
\end{tabular*}
\\
         $^a$ Fit using full data points up to 35000 cm$^{-1}$  \\
         $^b$ Fit using data points up to 10000 cm$^{-1}$\\
         $^c$ Fit using data points up to 5000 cm$^{-1}$\\
\caption{Comparison of Harmonic Frequencies (in cm$^{-1}$) between V$_{LL \rightarrow CC}$ PES computed using the force field and corresponding ab initio (CCSD(T)-F12a/aug-cc-pVDZ) for $syn$-ethanol.}
\label{tab:ff_freq_syn}
\end{table}

\end{document}